\documentclass[pra,tightenlines,superscriptaddress,eqsecnum,12pt]{revtex4}

\usepackage{graphicx}
\usepackage{mathrsfs}

\begin{document}

\def\bfk{{\bf k}}
\def\ket#1{|#1\rangle}
\def\bra#1{\langle#1|}
\def\tr{{\rm tr}}
\def\expect#1{{\langle#1\rangle}}
\def\inner#1#2{{\langle#1|#2\rangle}}
\def\outer#1#2{{|#1\rangle\langle#2|}}
\def\Hop{{\hat H}}
\def\Oop{{\hat O}}
\def\Odag{{\hat O}^\dagger}
\def\aop{{\hat a}}
\def\adag{{\hat a}^\dagger}
\def\bop{{\hat b}}
\def\bdag{{\hat b}^\dagger}
\def\cop{{\hat c}}
\def\cdag{{\hat c}^\dagger}
\def\Aop{{\hat A}}
\def\Adag{{\hat A}^\dagger}
\def\Bop{{\hat B}}
\def\Bdag{{\hat B}^\dagger}
\def\Cop{{\hat C}}
\def\Cdag{{\hat C}^\dagger}
\def\Lop{{\hat L}}
\def\Ldag{{\hat L}^\dagger}
\def\Sop{{\hat S}}
\def\Sdag{{\hat S}^\dagger}
\def\Uop{{\hat U}}
\def\Udag{{\hat U}^\dagger}
\def\proj{{\hat {\cal P}}}
\def\id{{\hat 1}}
\def\sx{{\hat \sigma}_x}
\def\sy{{\hat \sigma}_y}
\def\sz{{\hat \sigma}_z}
\def\r{\hat\rho}
\def\rtot{\hat\rho_{\rm total}}
\def\Pke{\hat P_{\bf k}^{\cal E}}
\def\Pre{\hat P_r^{\cal E}}
\def\tre{{\rm tr}_{\cal E}}
\def\Er{\hat E_r}
\def\Imin{I_{\rm min}}
\def\Dsbar{\Delta\bar H}
\def\Hbar{\bar H}
\def\Ibar{\bar I}
\def\Ds{\Delta H}
\def\cN{{\cal N}}
\def\sN{\mathscr{N}}
\def\cD{{\cal D}}
\def\ccD{{d}}  
\def\cS{{\cal S}}
\def\cE{{\cal E}}
\def\cH{{\cal H}}
\def\cB{{\cal B}}
\def\cK{{\cal K}}
\def\cV{{\cal V}}
\def\ND{\cN_\ccD(\phi)}
\def\NDb{\cN_\ccD(\phi_{\,b})}
\def\NDc{\cN_\ccD(\phi_c)}
\def\NDd{\cN_\ccD(\phi_d)}
\def\HD{H_\ccD(\phi)}
\def\HNmax{H_{\cN_V(\phi)}(\phi)}
\def\Sall{\cS_{2\ccD-1}}
\def\Splus{\cS_{2n-1}}
\def\Sminus{\cS_{2m-1}}
\def\Szero{\cS_{2(\ccD-n-m)-1}}
\def\rs{\hspace{-1pt}s\hspace{1pt}}  
\def\d{\mathrm{d}}

\newcommand{\one}{\hat{\mbox{\tt 1}\hspace{-0.057 in}\mbox{\tt l}}}

\title{Hypersensitivity and chaos signatures\\ in the quantum baker's maps}

\author{A. J. Scott}
\email{ascott@qis.ucalgary.ca}
\affiliation{Department of Physics and Astronomy,
University of New Mexico, Albuquerque, NM~87131-1156, USA}
\affiliation{Institute for Quantum Information Science,
University of Calgary, Calgary, Alberta T2N~1N4, Canada}

\author{Todd A. Brun}
\email{tbrun@usc.edu}
\affiliation{Communication Sciences Institute,
University of Southern California, Los Angeles, CA 90089-2565, USA}

\author{Carlton M. Caves}
\email{caves@info.phys.unm.edu}
\affiliation{Department of Physics and Astronomy,
University of New Mexico, Albuquerque, NM~87131-1156, USA}

\author{R\"udiger Schack}
\email{r.schack@rhul.ac.uk}
\affiliation{Department of Mathematics, Royal Holloway, University of
London, Egham, Surrey TW20 0EX, UK}

\date{18 September 2006}

\begin{abstract}
  \vspace{.1in} Classical chaotic systems are distinguished by their
  sensitive dependence on initial conditions.  The absence of this
  property in quantum systems has lead to a number of proposals for
  perturbation-based characterizations of quantum chaos, including linear
  growth of entropy, exponential decay of fidelity, and hypersensitivity
  to perturbation.  All of these accurately predict chaos in the classical
  limit, but it is not clear that they behave the same far from the
  classical realm.  We investigate the dynamics of a family of
  quantizations of the baker's map, which range from a highly entangling
  unitary transformation to an essentially trivial shift map.  Linear
  entropy growth and fidelity decay are exhibited by this entire family of
  maps, but hypersensitivity distinguishes between the simple dynamics of
  the trivial shift map and the more complicated dynamics of the other
  quantizations.  This conclusion is supported by an analytical argument
  for short times and numerical evidence at later times.
\end{abstract}


\maketitle

\section{Introduction}

A full characterization of quantum chaos is an elusive matter. Classical
chaotic systems are distinguished by their exponential sensitivity to
initial conditions. Quantified in terms of Lyapunov exponents, this
characterization is the key ingredient in any definition of classical
chaos. The linearity of quantum mechanics prohibits such sensitivity to
initial conditions, thus obstructing any straightforward extension of the
classical definition of chaos to quantum systems. The standard fix is to
categorize as ``chaotic'' those quantum systems that are chaotic in a
classical limit. These systems are not strictly chaotic by the classical
definition---they are quasiperiodic---but they have properties, involving
the spectrum of energy eigenvalues and the behavior of energy eigenstates,
that are distinctly different from that for quantizations of classically
regular systems.

So far, however, little agreement has been reached on a characterization
of quantum chaos that does not make reference to a classical limit.  The
classical approach of looking at small perturbations of the initial state
fails due to the unitarity of quantum dynamics.  Attempts at a direct
dynamical characterization of quantum chaos, which applies even in the
hard quantum regime, thus generally look at the effects of small
perturbations of the dynamics.  A number of perturbation-based criteria
(or signatures) have been proposed, including {\it linear growth of
entropy\/} in the presence of stochastic
perturbations~\cite{Zurek1994a,Zurek1995a}; {\it exponential decay of
fidelity\/} between states that evolve under two close, but distinct
unitary transformations~\cite{Peres1991b,Peres1993a}; and {\it
hypersensitivity to perturbation}, which considers stochastic
perturbations and compares the amount of information known about the
perturbation to the resulting reduction in the system
entropy~\cite{Caves1993b,Schack1996a,Schack1996b,Caves1997}, a relation
called the {\it information-entropy trade-off}.

Typically, a quantum system whose classical limit is chaotic exhibits all
three of these signatures.  They are clearly inequivalent, however. For
instance, hypersensitivity to perturbation can be viewed as a measure of
how fast, how widely, and how randomly the set of all possible perturbed
states spreads through Hilbert space. The other two criteria, though they
report on how widely the perturbed vectors are dispersed through Hilbert
space, are not sensitive to the exact way in which Hilbert space is
explored by the perturbed dynamics. It is therefore conceivable that there
are quantum systems that are chaotic with respect to one criterion, but
regular with respect to another. In this paper we compare the three
perturbation-based criteria for a family of quantizations of the quantum
baker's~map.

The classical baker's map~\cite{Lichtenberg1983} is a well-known toy
mapping whose study has led to many insights in the field of classical
chaos by demonstrating essential features of nonlinear dynamics.  It maps
the unit square, which can be thought of as a toroidal phase space, onto
itself in an area-preserving way.  Interest in the baker's map stems from
its straightforward formulation in terms of a Bernoulli shift on binary
sequences. It seems natural to consider a quantum version of the baker's
map for the investigation of quantum chaos. There is, however, no unique
procedure for quantizing a classical map; hence, different quantum maps
correspond to the same classical baker's transformation in the classical
limit.  The family of quantizations~\cite{Schack2000a} used in the present
paper is based on the $2^N$-dimensional Hilbert space of $N$ qubits.  This
qubit structure provides a connection to the binary representation of the
classical baker's map.

The paper is organized as follows. In Sec.~\ref{sec:criteria} we introduce
our mathematical notation and give precise definitions of the three
perturbation-based chaos criteria.  Section~\ref{sec:bakersmap} reviews a
family of quantizations for the quantum baker's map.  These range from a
highly entangling unitary transformation to an essentially trivial shift
map.  In Sec.~\ref{sec:extremalbakersmap} we give simple analytical
results concerning the three criteria for the trivial shift map; these
results show that the trivial shift map exhibits linear entropy growth and
exponential fidelity decay and suggest that it does not display
hypersensitivity to perturbation.  Section~\ref{sec:numerical} presents
numerical calculations for the entire family of baker's maps.  These
calculations show that, unlike the other two criteria, hypersensitivity to
perturbation differentiates between the different quantizations.  In
Appendix~\ref{app:tradeoff}, we formulate a simple model for the form of
the information-entropy trade-off in the case of vectors distributed
randomly on Hilbert space.  The model serves as a foil for interpreting
the results of our numerical work on hypersensitivity.
Appendix~\ref{app:entropy} derives the von Neumann entropy of an ensemble
of vectors that populate half of Hilbert space uniformly; this result is
used to bound the information-entropy trade-off in the case that the
amount of information about the perturbation is one bit.  Finally, in
Sec.~\ref{sec:conclusion} we discuss our results.

\section{Criteria for quantum chaos} \label{sec:criteria}

\subsection{Hypersensitivity to perturbation}

\subsubsection{Definition of hypersensitivity}
\label{sec:hyperdef}

In the most general setting, hypersensitivity to perturbation can be
defined as follows~\cite{Soklakov2000b}.  Consider a system with
Hilbert space ${\cS}$, evolving under some unitary evolution and, in
addition, interacting with an environment with Hilbert space ${\cE}$.
Let $D$ and $D_{\cE}$ denote the Hilbert-space dimensions of the system
and environment, respectively. Initially, the joint state of the system and
environment is assumed to be a product state; i.e., initially there is
no correlation. After a time $t$, the joint state of the system and
environment is a density operator on ${\cS}\otimes{\cE}$, which
we denote by $\rtot$. The state of the system at time $t$, $\r$, is
obtained by tracing out the environment,
\begin{equation}   \label{eq:trRtot}
\r=\tre(\rtot) \;.
\end{equation}
The von Neumann entropy of the system at time $t$ is
\begin{equation}
H_{\cS} = -\tr(\r\log\r) \;.
\label{eq:vonNeumann}
\end{equation}
We measure entropy in bits (i.e. we take $\log\equiv\log_2$).

Now assume that an arbitrary measurement is performed on the
environment.  The most general measurement~\cite{Kraus1983} is
described by a positive-operator-valued measure (POVM), $\{\Er\}$,
where the $\Er$ are positive operators acting on the environment and
satisfying the completeness condition
\begin{equation}
\sum_r\Er=\one_{\cE}=\mbox{(environment identity operator).}
\end{equation}
The probability of obtaining the measurement outcome $r$ is given by
\begin{equation}
p_r= \tr\!\left(\rtot(\one_{\cS}\otimes\Er)\right)\;,
\end{equation}
where $\one_{\cS}$ is the identity operator on the system.  The
system state after a measurement that yields the outcome $r$ is
\begin{equation}
\r_r=
{\tre\!\left( \rtot(\one_{\cS}\otimes\Er) \right)\over p_r}\;.
\end{equation}
We define the system entropy conditional on the outcome $r$,
\begin{equation}
H_r = -\tr(\r_r\log\r_r) \;,
\end{equation}
the average conditional entropy,
\begin{equation}
\Hbar = \sum_r p_r H_r \;,
\end{equation}
and the average entropy decrease due to the measurement,
$\Dsbar=H_{\cS}-\Hbar$. Furthermore, we define the {\em average
information},
\begin{equation}
\Ibar = -\sum_r p_r \log p_r \;.
\end{equation}
The quantity $\Ibar$ is within 1 bit of the minimum average
algorithmic information needed to specify the measurement outcome $r$
\cite{Schack1997a}.

Now assume we want to perform a measurement that reduces the average
conditional system entropy below some given target value, $H$.
We define the quantity
\begin{equation}
\Imin(H) = \inf \Ibar \;,
\label{eq:defimin}
\end{equation}
where the infimum is taken over all POVMs $\{\Er\}$ such that $\Hbar\le
H$. The function $\Imin(H)$ expresses what we call the {\em
information-entropy trade-off}\/; it can be interpreted as the minimum
information about the perturbing environment needed to keep the average
system entropy below the target value $H$.  We say the system is {\em
hypersensitive to perturbation\/} if this information is large compared to
the purchased entropy reduction $\Ds=H_{\cS}-H$, i.e.,
\begin{equation}
\frac{\Imin(H)}{\Ds}\gg 1\;,
\end{equation}
in the region of small enough entropy reductions that this ratio reports
on the system dynamics rather than on the multiplicity of possible
perturbations.  We characterize this region more precisely in the next
subsection.

For the analysis of the present paper, we specialize to the case where
initially the system is in a pure state, $\ket{\psi_0}$, and the
unitary system evolution is given by a quantum map $\Bop$. In the
absence of any interaction with the environment, the system state after
$t$ iterations (or time steps) is $\Bop^t\ket{\psi_0}$.  In addition,
we assume that the effect of the environment is equivalent to a
stochastic perturbation. At each time step, a perturbation is chosen
randomly from a set of unitary maps, $\{U_k:k\in{\cK}\}$, where
$\cK$ is some index set.

The joint system-environment density operator after $t$ iterations is
then given by
\begin{equation}   \label{eq:jointStates2}
\rtot=\sum_{\bfk\in{\cK}^t} p_\bfk \ket{\psi_\bfk}\bra{\psi_\bfk} \otimes \Pke  \;,
\end{equation}
where
\begin{equation}
\ket{\psi_\bfk} =
\hat U_{k_t}\Bop\hat U_{k_{t-1}}\Bop\cdots \hat U_{k_1}\Bop\ket{\psi_0}
\end{equation}
is the endpoint of a stochastic trajectory labeled by
$\bfk=(k_1,\ldots,k_t)$, $p_\bfk$ is the probability of $\bfk$, and the
operators $\Pke$ are one-dimensional, orthogonal environment
projectors.  The perturbation histories $\bfk=(k_1,\ldots,k_t)$ are
thus recorded in the environment in the form of the orthogonal projectors $\Pke$.
The reduced density operator for the system is given by
\begin{equation}
\r = \tr_{\cE}(\rtot) = \sum_{\bfk\in{\cK}^t} p_\bfk \ket{\psi_\bfk}\bra{\psi_\bfk}\;.
\end{equation}
In the numerical analysis in Sec.~\ref{sec:numerical}, we let the
environment be a series of qubits, which interact sequentially with the
system; hence the perturbation at each time step is a binary
perturbation, $\cK=\{0,1\}$.

It is in general very difficult to determine the function $\Imin(H)$
because it is generally impossible to do the required optimization
over all POVMs.  In the numerical results reported in
Sec.~\ref{sec:numerical}, we restrict the optimization to POVMs of
the form
\begin{equation}
\Er=\sum_{\bfk\in K_r}\Pke \;,
\label{eq:groupingPOVM}
\end{equation}
where the subsets $K_r\subset{\cK}^t$ are nonoverlapping subsets of the
perturbation histories.  Such POVMs can be regarded as sampling a
coarse-grained version of the perturbation histories. For the special form
of $\rtot$ considered here, it seems reasonable that ensembles which are
optimal with respect to this class of measurements are also, to a good
approximation, optimal with respect to the class of all possible
environment measurements. We have not, however, been able to prove this
statement rigorously.

The measurements~(\ref{eq:groupingPOVM}) correspond to forming groups of
system vectors $\ket{\psi_\bfk}$.  Assuming that all perturbations are
equally likely, i.e., $p_\bfk=1/\cN$ for all $\bfk$, where $\cN$ is the
number of perturbation histories or system vectors $\ket{\psi_\bfk}$,
the probability of obtaining outcome $r$ is given by
\begin{equation}
p_r= |K_r|/\cN \;,
\end{equation}
and the system state after a measurement that yields outcome $r$ is the
average of the grouped vectors,
\begin{equation}
\r_r={\tre\!\left( \rtot(\one_{\cS}\otimes\Er) \right)\over p_r}={1\over|K_r|}\sum_{\bfk\in K_r}
\ket{\psi_\bfk}\bra{\psi_\bfk} \;.
\end{equation}
We use this simplified framework in the discussion in Secs.~IV and~V
below.

Even this simplified framework is not sufficient to make the problem
tractable for numerical purposes, because the number of vectors, $\cN$,
increases exponentially with the number of time steps, rapidly making
it impossible to search over all possible ways of grouping the vectors
$\ket{\psi_\bfk}$.  To get around this, we employ efficient algorithms
for grouping the vectors, which are plausibly able to find optimal or
near-optimal groupings.  For the numerical results reported in Sec.~\ref{sec:numerical},
we first use a particularly simple, but intuitive grouping algorithm devised to
clarify the procedure and then take a general approach based on
genetic algorithms.  We have compared the results
obtained using these algorithms with those obtained using other grouping algorithms,
some of which have been used previously~\cite{Schack1996b}, and found that the
current algorithms are generally superior for the vectors
generated by the perturbed baker's map.

\subsubsection{Quantitative measure of hypersensitivity}
\label{subsubsec:quanthyper}

Hypersensitivity to perturbation tests how fast and how fully the state of
the perturbed system explores the system Hilbert space. To see how this is
quantified by the information-entropy trade-off, we consider the trade-off
relation for vectors that are distributed randomly in Hilbert space. Such
a relation was formulated in~\cite{Schack1996b,Soklakov2000b}, using a
model that groups the random vectors into spheres of uniform radius
(measured by Hilbert-space angle) on projective Hilbert
space~\cite{Schack1994a}. We refine this model and its trade-off relation
in Appendix~\ref{app:tradeoff}.  The main result is that for $\cN$ vectors
distributed randomly in $\ccD$ Hilbert-space dimensions, the
information-entropy trade-off, written in inverse form, is approximated by
\begin{equation}
H=\cases{
  \log\cN-\Imin\;,
        &$\log\cN\ge \Imin\ge\log\cN-\log\ccD$,\cr
    \log\ccD-\displaystyle{{1\over\ccD}}
    \Bigl((1+\Imin\ln2)\log(1+\Imin\ln2)-\Imin\Bigr)\;,&$\log\cN-\log\ccD\ge\Imin\ge1$.}
\label{eq:tradeoff}
\end{equation}
This expression assumes that $\ccD$ is large and that the number of random
vectors, though large in the sense that $\cN\gg\ccD$, satisfies $\cN\ll
2^{\ccD}$, a situation we refer to as a \textit{sparse\/} collection of
vectors.  Examples of the information-entropy trade-off are shown in
Fig.~\ref{fig3} (Sec.~\ref{sec:numerical}), in which the upper solid curve
closely resembles the exact situation for random vectors in 32 dimensions;
notice that $\Imin$ follows $H$ in a linear fashion before dropping
quickly at a ``knee'' close to the maximum entropy.  This agrees with our
approximation~(\ref{eq:tradeoff}) for random vectors, which is shown as
the upper dotted line and the rightmost dotted curve in Fig.~\ref{fig3}.

Several features of the sphere-grouping trade-off~(\ref{eq:tradeoff})
deserve discussion.  The number of spheres, $2^{\Imin}$, gives the number
of vectors per group $\cN_V=\cN2^{-\Imin}$.  The knee at
$\Imin=\log\cN-\log\ccD$ thus corresponds to $\cN_V=\ccD$.  For
$\Imin>\log\cN-\log\ccD$, i.e., $\cN_V<\ccD$, the number of vectors in
each sphere is insufficient to explore all the Hilbert-space dimensions.
This gives a linear dependence on $\Imin$, with slope $-1$ and intercept
$\log\cN$.  In the context of a stochastically perturbed map, where $\cN$
is the number of perturbation histories, this part of the trade-off
relation tells us about the multiplicity of the perturbation instead of
about the dynamics of the map. In contrast, for $\Imin<\log\cN-\log\ccD$,
i.e., $\cN_V>\ccD$, where the number of vectors in each sphere is large
enough to explore all Hilbert-space dimensions, the information-entropy
trade-off becomes independent of $\cN$.  {\em It is this part of the
trade-off relation, beyond the knee in the information-entropy trade-off,
that tells us about hypersensitivity to perturbation in the system
dynamics}.  Notice that we need $\cN\gg\ccD$ to investigate this region,
but we do {\it not\/} need $\cN$ so large that a random collection of
vectors would sample generic vectors, which requires at least
$\cN\sim2^{\ccD}$ vectors, i.e., what we call a dense collection.  Our
stochastic perturbation need only produce a sparse collection of vectors
to see evidence of hypersensitivity; we can say that the vectors in such a
sparse collection are pseudo-random instead of random~\cite{Emerson2003}.

Projective Hilbert space can never be partitioned exactly into
spheres of uniform radius.  This has little effect when the
spheres are tiny and numerous, but it becomes a problem when there
are just a few spheres and prompts us to treat the sphere-grouping
trade-off relation with caution in this situation.  In particular,
for the case of just two groups, i.e., $\Imin=1$, a better
method for grouping random vectors is to partition
projective Hilbert space into two equal volumes defined by the
closeness to two orthogonal subspaces of dimension $\ccD/2$.
The resulting trade-off for $\Imin=1$ is analyzed in
Appendix~\ref{app:entropy} and summarized below.  We abandon the
trade-off relation~(\ref{eq:tradeoff}) entirely for $\Imin<1$
because a grouping into spheres of uniform radius makes no sense
when there are fewer than two spheres.

We are now in a position to introduce a quantitative measure of
a map's hypersensitivity to perturbation. For this purpose we
introduce the quantity
\begin{equation}
\label{eq:iquantity}
\rs\equiv\frac{1}{H_{\cS}-H(\Imin=1)} \;;
\end{equation}
$1/\rs$ is the reduction in system entropy purchased by gathering one
(optimal) bit of information about the environment. In a Hilbert-space
context, $\rs$ is an indicator of the randomness in a collection of
vectors.  It is a considerably more informative indicator than the
entropy.  For example, the members of an orthonormal basis together
achieve maximal entropy, yet a grouping of these vectors into two equally
sized groups gives $s=1$, independent of the dimension.  In contrast, for
vectors distributed randomly across Hilbert space, the sphere-grouping
trade-off relation~(\ref{eq:tradeoff}) gives
\begin{equation}
\rs=\frac{1}{\log\ccD-H(\Imin=1)}={\ccD\over(1+\ln2)\log(1+\ln2)-1}\approx 3.5\,\ccD\;.
\label{eq:sspheres}
\end{equation}
While~$\rs$ provides a signature of randomness, it is the change in~$\rs$
with time, as applied to the perturbed system vectors, which indicates the
degree to which a system is hypersensitive to perturbation.  A rapid
increase in~$\rs$ over time has been proposed as a criterion of chaos for
both classical and quantum systems~\cite{Caves1993b}.  In particular,
if~$\rs$ increases exponentially with time, we say that the system
exhibits \textit{exponential hypersensitivity to
perturbation\/}~\cite{Schack1996a,Caves1997}.

A detailed analysis~\cite{Schack1996a} of stochastic perturbations of
classically chaotic maps described by a symbolic dynamics shows that for
such systems, $\rs$ is indeed a measure of the phase-space stretching and
folding characteristic of chaotic dynamics.  Specifically, $\rs$ grows as
$2^{Kt}$, where $K$ is the Kolmogorov-Sinai entropy of the
dynamics~\cite{Alekseev1981}, showing that these systems do display
exponential hypersensitivity to perturbation and that exponential
hypersensitivity is equivalent to the standard characterization of
classical chaos via the the Kolmogorov-Sinai~entropy, which in turn is
equivalent to characterization in terms of Lyapunov exponents.

For quantum systems, $\rs$ is a measure of how fast and how fully the
state of a perturbed system explores the system Hilbert space.  An
exponential increase in~$\rs$ indicates both that the number of
dimensions, $\ccD$, explored by the perturbed vectors grows exponentially
and that the vectors populate the explored dimensions randomly.  Thus $s$
provides a direct dynamical characterization of quantum chaotic dynamics,
a characterization that is analogous to the characterization of
classical chaos in terms of sensitivity to initial conditions.  The reason
that hypersensitivity to perturbation goes beyond the Zurek-Paz chaos
criterion of linear entropy increase under stochastic
perturbations~\cite{Zurek1994a,Zurek1995a} is clear: a linear entropy
increase indicates that the perturbed vectors explore an exponentially
increasing number of dimensions, but is silent on whether those dimensions
are explored randomly.

A related parameter for characterizing hypersensitivity is the slope of
the information-entropy trade-off, $|d\Imin/dH|$, evaluated at $\Imin=0$
(i.e., $H=H_{\cS}$) or perhaps at $\Imin=1$.  Both the classical analysis
in~\cite{Schack1996a} and the analysis of Appendix~\ref{app:tradeoff}
prompt us to shy away from using the slope evaluated at $\Imin=0$, since
there are uncertainties about the behavior of the slope for very small values
of $\Imin$.  Moreover, the slope evaluated at $\Imin=1$ seems to have no
advantages over the parameter~$\rs$.  Thus, in this paper, we calculate
numerically information-entropy trade-offs for the perturbed quantum
baker's maps, and from these we determine the time evolution of the
hypersensitivity parameter~$\rs$, preferring it to the more problematic
use of the slope.

Having settled on~$\rs$ as our signature of hypersensitivity, we can
formulate a better information-entropy trade-off for random vectors when
$\Imin=1$, i.e., for the case of two groups.  An optimal way of grouping a
sufficiently dense collection of random vectors, analyzed in
Appendix~\ref{app:entropy}, is then the following: choose two orthogonal
subspaces, each of dimension $\ccD/2$, and partition projective Hilbert
space into two equal volumes defined by the distance in Hilbert-space
angle to these subspaces.  The entropy of each partition is
\begin{equation}
H={\ccD\over2}\Bigl(-\lambda_+\log\lambda_+ -\lambda_-\log\lambda_-\Bigr)
\approx\log\ccD-{1\over\pi\ccD\ln2}\label{eq:ppentropy}
\end{equation}
[cf.~Eqs.~(\ref{eq:HnD}) and (\ref{eq:HnDapprox}) with $n=\ccD/2$],
where
\begin{equation}
\lambda_\pm=
{1\over\ccD}\left(1\pm{\ccD!\over2^\ccD[(\ccD/2)!]^2}\right)
\approx{1\over\ccD}\left(1\pm{\sqrt{2\over\pi\ccD}}\right)
\end{equation}
[cf.~Eqs.~(\ref{eq:lambdapm}) and (\ref{eq:lambdapmapprox}) with
$n=\ccD/2$].  The approximate expressions on the right hold for large
$\ccD$ and give
\begin{equation}
\rs={1\over\log\ccD-H}=\pi\ccD\ln2\approx 2.2\,\ccD\;.
\label{eq:s}
\end{equation}
The coefficient 2.2, smaller than the 3.5 of Eq.~(\ref{eq:sspheres}),
indicates that this is a better way to partition random vectors into
two groups. This value of~$\rs$ represents an approximate upper bound for any
collection of vectors in Hilbert space.

Another scenario that is important for the current study occurs when the
perturbed vectors are restricted to product states of $N$ qubits. Random
product vectors can be grouped into the two groups corresponding to
$\Imin=1$ by partitioning the projective Hilbert space of one of the
qubits into two equal volumes, just as above.  The entropy of each
partition for this qubit is $2-(3/4)\log 3$ [Eq.~(\ref{eq:ppentropy}) with
$\ccD=2$]. Thus for all the qubits, the entropy of each partition is
\begin{equation}
H=N-\frac{3}{4}\log 3+1\;,
\end{equation}
which gives
\begin{equation}
\rs={1\over N-H}=\frac{4}{3\log 3 -4}\approx 5.3\;,
\label{eq:s2}
\end{equation}
independent of $D=2^N$. This value represents a rather restrictive
approximate upper bound on~$\rs$ for product vectors.

Suppose that for random product vectors, we partition the projective
Hilbert spaces of $j$ constituent qubits into two equal volumes, thus
using $j=\Imin$ bits of information to purchase a reduction of the entropy to
\begin{equation}
H(\Imin)=N-\left(\frac{3}{4}\log 3-1\right)\Imin
\approx N-\Imin/5.3\;,
\label{eq:tradeoff2}
\end{equation}
for $N\ge\Imin\ge0$. This information-entropy trade-off, which, unlike
Eq.~(\ref{eq:tradeoff}), is linear near the maximal entropy, is plotted as
the lower dotted line in Fig.~\ref{fig3} (Sec.~\ref{sec:numerical}).  It
shows that nonentangling quantum maps are not hypersensitive to
perturbation.

\subsection{Other perturbation-based criteria for quantum chaos}

Fidelity decay as a criterion for quantum chaos was introduced by Peres
\cite{Peres1991b,Peres1993a} (see
also~\cite{Emerson2002,Weinstein2005,Gorin2006} and references therein).
One compares the unitary evolution of an initial state $\ket{\psi_0}$
under the action of a quantum map $\Bop$ with the evolution of the same
initial state under the action of a modified map, $\Bop'=\hat U\Bop$,
where the unitary map $\hat U$ is close to the identity operator.
According to this criterion, a quantum map is chaotic if the fidelity,
\begin{equation}
F(t)  = \bigl|\bra{\psi_0}\bigl(\Bdag\hat U^\dagger\bigr)^t\Bop^t\ket{\psi_0}\bigr|^2\;,
\end{equation}
decreases exponentially with the number of iterations at short times. In
contrast to the criterion of hypersensitivity to perturbation, where the
effects of a stochastic perturbation are analyzed, fidelity decay focuses
on just two perturbation histories, corresponding to the unperturbed
evolution and to a modified evolution where the same perturbation operator
$\hat U$ is applied at each time step.

Linear entropy increase as a chaos criterion was introduced by Zurek and
Paz~\cite{Zurek1994a,Zurek1995a}.  According to this criterion, a quantum
map is chaotic if the entropy~(\ref{eq:vonNeumann}) of the reduced system
density operator~(\ref{eq:trRtot}) increases linearly with the number of
iterations at short times.  As we have already discussed, a linear entropy
increase is essential for exponential hypersensitivity to perturbation,
but it is not the whole story.

\section{Quantum baker's maps} \label{sec:bakersmap}

The classical baker's map is a standard example of chaotic dynamics
\cite{Lichtenberg1983}.  It is a symplectic map of the unit square onto
itself, defined through the equations
\begin{eqnarray}
q_{n+1} &=& 2 q_n - \lfloor 2q_n \rfloor\;, \label{eq:classbake1}\\
p_{n+1} &=& \left(p_n + \lfloor2 q_n \rfloor \right)/2\;, \label{eq:classbake2}
\end{eqnarray}
where $q,p \in [0,1)$, $\lfloor x \rfloor$ is the integer part of $x$,
and $n$ denotes the $n$th iteration of the map. Geometrically, the map
stretches the unit square by a factor of two in the $q$ direction,
squeezes by a factor of a half in the $p$ direction, and then stacks
the right half onto the left.

Interest in the baker's map stems from its straightforward
symbolic-dynamical characterization in terms of a Bernoulli shift on
binary sequences. If each point of the unit square is identified
through its binary representation, $q = 0\!\cdot\!s_1 s_2 \ldots =
\sum_{k=1}^\infty s_k 2^{-k}$ and $p= 0\!\cdot\! s_0 s_{-1} \ldots =
\sum_{k=0}^\infty s_{-k} 2^{-k-1}$ ($s_i\in\{0,1\}$), with a
bi-infinite symbolic string
\begin{equation}
s = \underbrace{\ldots s_{-2} s_{-1} s_0}_p \bullet
\underbrace{s_1 s_2 s_3 \ldots}_q \;\,,
\label{eq:symbseq}
\end{equation}
then the action of the baker's map is to shift the position of the dot
by one digit to the right,
\begin{equation}
s\rightarrow s'
= \underbrace{\ldots s_{-2} s_{-1} s_0 s_1}_{p'} \bullet
\underbrace{s_2 s_3 \ldots}_{q'} \;\,.
\end{equation}

It seems natural to consider a quantum version of the baker's map for
the investigation of quantum chaos. There is, however, no unique
procedure for quantizing a classical map: different quantum maps can
lead to the same classical baker's transformation.

To construct a quantum baker's map, we work in a $D$-dimensional
Hilbert space, ${\cH}_D$, spanned by either the position states
$\ket{q_j}$, with eigenvalues $q_j=(j+1/2)/D$, or the momentum states
$\ket{p_k}$, with eigenvalues $p_k =(k+1/2)/D$ ($j,k=0,\ldots,D-1$).
The constants of $1/2$ determine the type of periodicity assumed for
the position and momentum states, in this case, $\ket{q_{j+D}} =
-\ket{q_j}$, $\ket{p_{k+D}} = -\ket{p_k}$, and thus identify ${\cH}_D$
with a toroidal phase space with antiperiodic boundary conditions. The
vectors of each basis are orthonormal,
$\inner{q_j}{q_k}=\inner{p_j}{p_k}=\delta_{jk}$, and the two bases are
related via the discrete Fourier transform $\hat F_D$,
\begin{equation}
\bra{q_j}\hat{F}_D\ket{q_k}\equiv\inner{q_j}{p_k}=
\frac{1}{\sqrt{D}}\,e^{iq_jp_k/\hbar}\;.
\end{equation}
For consistency of units, we must have $2 \pi \hbar D=1$.

The first work on a quantum baker's map was done by Balazs and Voros
\cite{Balazs1989}.  Assuming an even-dimensional Hilbert space with
periodic boundary conditions, they defined a quantum baker's map in terms
of a unitary operator $\Bop$ that executes a single iteration of the map.
Saraceno~\cite{Saraceno1990} later improved certain symmetry
characteristics of this quantum baker's map by using antiperiodic boundary
conditions as described above. To define the Balazs-Voros-Saraceno unitary
operator in our notation, imagine that the even-dimensional Hilbert space
is a tensor product of a qubit space and the space of a
($D/2$)-dimensional system. Writing $j=x(D/2)+j'$, $x\in\{0,1\}$, we can
write the position eigenstates as $\ket{q_j}=\ket{x}\otimes\ket{j'}$,
where the states $\ket{x}$ make up the standard basis for the qubit, and
the states $\ket{j'}$ are a basis for the ($D/2$)-dimensional system.  The
state of the qubit thus determines whether the position eigenstate lies in
the left or right half of the unit square.  The Balazs-Voros-Saraceno
quantum baker's map is defined by
\begin{equation}
\Bop=\hat{F}_D\circ\Bigl(\one_2\otimes\hat F_{D/2}^{-1}\Bigr)\;,
\label{eq:BV}
\end{equation}
where $\one_2$ is the unit operator for the qubit, and
$\hat{F}_{D/2}$ is the discrete Fourier transform on the
($D/2$)-dimensional system.  The unitary $\Bop$ does separate
inverse Fourier transforms on the left and right halves of the unit
square, followed by a full Fourier transform.

For dimensions $D=2^N$, an entire class of quantum baker's maps
can be defined in analogy with the symbolic dynamics for the classical baker's
map~\cite{Schack2000a}. In this case,
we can model our Hilbert space as the tensor-product space of
$N$ qubits, and the position states can be defined as product states
for the qubits in the standard basis, i.e.,
\begin{equation}
\ket{q_j}=
\ket{x_1}\otimes\ket{x_2}\otimes\cdots\otimes\ket{x_N}\;,
\label{eq:qj}
\end{equation}
where $j$ has the binary expansion
\begin{equation}
j=x_1\ldots x_N\!\cdot\! 0=\sum_{l=1}^N x_l 2^{N-l}
\end{equation}
and $q_j=(j+1/2)/D=0\!\cdot\! x_1\ldots x_N1$.

To make the connection with the symbolic dynamics for the classical
baker's map, we proceed as follows.  The bi-infinite
strings~(\ref{eq:symbseq}) that specify points in the unit square
are replaced by sets of orthogonal quantum states created through
the use of a partial Fourier transform
\begin{equation}
\hat G_n\equiv\one_{2^n}\otimes\hat F_{2^{N-n}}\;,\qquad n=0,\ldots,N,
\end{equation}
where $\one_{2^n}$ is the unit operator on the first $n$ qubits and
$\hat F_{2^{N-n}}$ is the Fourier transform on the remaining qubits.
The partial Fourier transform thus transforms the $N-n$ least
significant qubits of a position state,
\begin{eqnarray}
\label{eq:partialfourier}
&& \hat{G}_n\,
\ket{x_1}\otimes\cdots\otimes\ket{x_n}\otimes
\ket{a_1}\otimes\cdots\otimes\ket{a_{N-n}}   \cr
&& \;\;\;\; =\,\ket{x_1}\otimes\cdots\otimes\ket{x_n}\otimes
\frac{1}{\sqrt{2^{N-n}}}\sum_{x_{n+1},\ldots,x_N}\ket{x_{n+1}}
\otimes\cdots\otimes\ket{x_N}e^{2\pi iax/2^{N-n}}\;,
\end{eqnarray}
where $a$ and $x$ are defined through the binary representations
$a=a_1\ldots a_{N-n}\!\cdot\! 1$ and $x=x_{n+1}\ldots x_N\!\cdot\!
1$. In the limiting cases, we have $\hat{G}_0=\hat{F}_{D}$ and
$\hat{G}_N=i\one$.  The analogy to the classical case is made clear
by introducing the following notation for the partially transformed
states:
\begin{equation}
\ket{\,a_{N-n}\ldots a_1\bullet x_1\ldots x_n}
\,\equiv\,\hat{G}_n\,\,\ket{x_1}\otimes\cdots\otimes\ket{x_n}\otimes
\ket{a_1}\otimes\cdots\otimes\ket{a_{N-n}}\;.
\label{eq:partialfourierstates}
\end{equation}
For each value of $n$, these states form an orthonormal basis and are
localized in both position and momentum.  The state $\ket{a_{N-n}\ldots
a_1\bullet x_1\ldots x_n}$ is strictly localized in a position region of
width $1/2^n$ centered at $0\!\cdot\! x_1 \ldots x_n 1$ and is roughly
localized in a momentum region of width $1/2^{N-n}$ centered at
$0\!\cdot\! a_1 \ldots a_{N-n} 1$.  In the notation of
Eq.~(\ref{eq:symbseq}), it is localized at the phase-space point $1a_{N-n}
\ldots a_1 \bullet x_1 \ldots x_n 1$. Notice that $\ket{a_{N}\ldots
a_1\bullet}= \hat{G}_0\,\ket{a_1}\otimes\cdots\otimes\ket{a_{N}}$ is a
momentum eigenstate and that $\ket{\!\bullet x_1\ldots x_N}=
\hat{G}_N\,\ket{x_1}\otimes\cdots\otimes\ket{x_N}=\,
i\ket{x_1}\otimes\cdots\otimes\ket{x_N}$ is a position eigenstate, the
factor of $i$ being a consequence of the antiperiodic boundary conditions.

Using this notation, a quantum baker's map on $N$ qubits is defined
for each value of $n=1,\ldots,N$ by the single-iteration unitary
operator~\cite{Schack2000a}
\begin{eqnarray}
\Bop_{N,n} &\equiv& \hat{G}^{\phantom{-1}}_{n-1} \circ \hat{S}_n \circ \hat{G}^{\,-1}_n \nonumber\\
&=&\sum_{x_1,\dots,x_n}\sum_{a_1,\dots,a_{N-n}}
\ket{\,a_{N-n}\dots a_1 x_1\bullet x_2\dots x_n} \bra{a_{N-n}\dots a_1
\bullet x_1 x_2\dots x_n}\;,
\label{eq:bakern}
\end{eqnarray}
where the shift operator $\hat{S}_n$ acts only on the first $n$
qubits, i.e., $\hat{S}_n\ket{x_1}\otimes\ket{x_2}\otimes\cdots\otimes
\ket{x_n}\otimes\ket{x_{n+1}}\otimes\cdots\otimes\ket{x_N}
=\ket{x_2}\otimes\cdots\otimes\ket{x_n}\otimes\ket{x_1}
\otimes\ket{x_{n+1}}\otimes\cdots\otimes\ket{x_N}$.  Notice that
since $\hat S_n$ commutes with $\hat G_n^{\,-1}$, we can put $\hat
B_{N,n}$ in the form
\begin{equation}
\Bop_{N,n}=\one_{2^{n-1}}\otimes
\Bigl(\hat F_{2^{N-n+1}}\circ(\one_2\otimes\hat F_{2^{N-n}}^{-1})\Bigr)
\circ\hat S_n\;.
\end{equation}
Since $\hat S_1$ is the unit operator, it is clear that $\Bop_{N,1}$ is
the Balazs-Voros-Saraceno quantum baker's map~(\ref{eq:BV}). It is worth
mentioning here that Ermann and Saraceno~\cite{Ermann2006} have recently
proposed and investigated an even larger family of quantum baker's maps,
which includes all of the above quantizations as members.  For the
purposes of this article, however, we need only consider $\Bop_{N,n}$.

We can also write
\begin{equation}
\Bop_{N,n}=\one_{2^{n-1}}\otimes\Bop_{N-n+1,1}\circ\hat S_n\;, \label{eq:Bshiftunit}
\end{equation}
which shows that the action of $\Bop_{N,n}$ is a shift of the $n$
leftmost qubits followed by an application of the Balazs-Voros-Saraceno
baker's map to the $N-n+1$ rightmost qubits.  At each iteration, the
shift map $\hat S_n$ does two things: it shifts the $n$th qubit, the
most significant qubit in position that was subject to the previous application
of $\Bop_{N-n+1,1}$, out of the region subject to the next
application of $\Bop_{N-n+1,1}$, and it shifts the most significant
qubit in position (first qubit) into the region of subsequent application of $\hat
B_{N-n+1,1}$.

\begin{figure}[t]
\includegraphics[width=14cm]{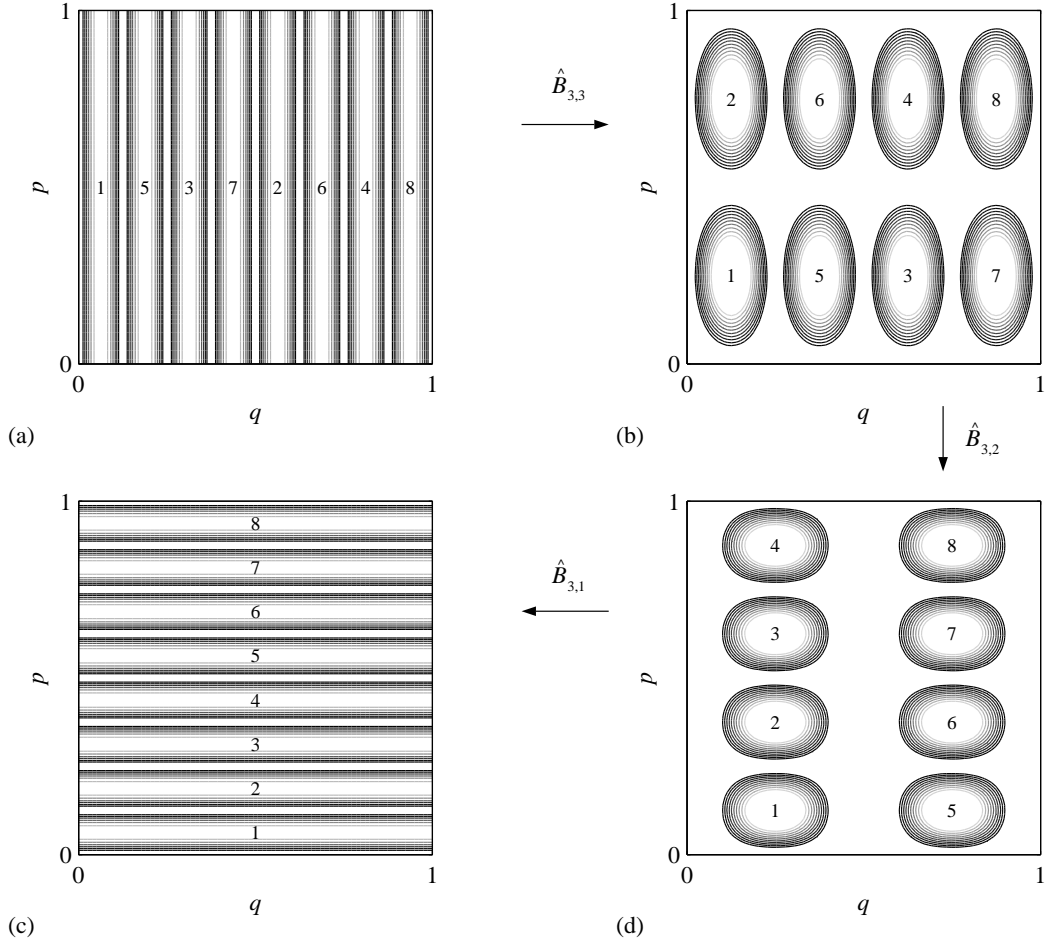}
\caption{Husimi function for each partially Fourier transformed state
(\ref{eq:partialfourierstates}) when $N=3$: (a) $n=3$, (b) $n=2$, (c)
$n=0$, and (d) $n=1$.  The action of the quantum baker's map $\Bop_{3,3}$
is to map the eight states in (a) to the eight states in (b), as shown by
the numbers labeling the states.  Similarly, $\hat B_{3,2}$ and $\hat
B_{3,1}$ map one set of partially Fourier transformed states to another,
as indicated by the arrows.  The map $\hat B_{3,1}$ is the
Balazs-Voros-Saraceno quantum baker's map.\label{fig1}}
\end{figure}

The quantum baker's map $\Bop_{N,n}$ takes a state localized at $1a_{N-n}
\ldots a_1 \bullet x_1 \ldots x_n 1$ to a state localized at $1a_{N-n}
\ldots a_1 x_1 \bullet x_2 \ldots x_n 1$.  The decrease in the number of
position bits and increase in momentum bits enforces a stretching and
squeezing of phase space in a manner resembling the classical baker's map.
In Fig.~\ref{fig1}(a), (b), (c), and (d), we plot the Husimi function
(defined as in~\cite{Tracy2002}) for the partially Fourier transformed
states~(\ref{eq:partialfourierstates}) when $N=3$, and $n=3$, 2, 0, and 1,
respectively.  The quantum baker's map is a one-to-one mapping of one
basis to another, as shown in the figure.

One useful representation of our quantum baker's maps, introduced in
\cite{Schack2000a}, starts from using standard techniques
\cite{Nielsen2000b} to write the partially transformed
states~(\ref{eq:partialfourier}) as product states:
\begin{eqnarray}
\ket{\,a_{N-n}\ldots a_1\bullet x_1\ldots x_n}&=&
e^{\pi i(0\cdot a_1\ldots a_{N-n}1)} \Biggl(\,\bigotimes_{k=1}^n\ket{x_k}\Biggr) \nonumber\\
&&\otimes\Biggl(\,\bigotimes_{k=n+1}^N{1\over\sqrt2}\Bigl(
\ket{0}+e^{2\pi i(0\cdot a_{N-k+1}\ldots a_{N-n}1)}\ket{1}\Bigr)
\Biggr)\;.
\end{eqnarray}
These input states are mapped by $\Bop_{N,n}$ to output states
\begin{eqnarray}
&\mbox{}& \ket{\,a_{N-n}\ldots a_1x_1\bullet x_2\ldots x_n}\,=\,
e^{\pi i(0\cdot x_1a_1\ldots a_{N-n}1)}
\Biggl(\,\bigotimes_{k=2}^n\ket{x_k}\Biggr) \cr
&\mbox{}& \;\;\;\otimes\Biggl(\,\bigotimes_{k=n+1}^N
{1\over\sqrt2}\Bigl(
\ket{0}+e^{2\pi i(0\cdot a_{N-k+1}\ldots a_{N-n}1)}\ket{1}\Bigr)
\Biggr)
\otimes
{1\over\sqrt2}\Bigl(
\ket{0}+e^{2\pi i(0\cdot x_1a_1\ldots a_{N-n}1)}\ket{1}\Bigr)
\;.\qquad
\end{eqnarray}
These forms show that the quantum baker's map $\Bop_{N,n}$ shifts the
states of all the qubits to the left, except the state of the leftmost
qubit.  The state $\ket{x_1}$ of the leftmost qubit can be thought of
as being shifted to the rightmost qubit, where it suffers a controlled
phase change that is determined by the state parameters
$a_1,\ldots,a_{N-n}$ of the original ``momentum qubits.''  The quantum
baker's map can thus be written as a shift map on a finite string of
qubits, followed by a controlled phase change on the least significant
qubit in position.  In~\cite{Soklakov2000a} this shift representation
was developed into a useful tool.  Using an approach based on coarse
graining in this representation, the classical limit of the quantum
baker's maps was investigated.

The classical limit for the above quantum baker's maps was also
investigated in~\cite{Tracy2002}, using an analysis based on the
limiting behavior of the coherent-state propagator of $\Bop_{N,n}$.
When $D=2^N\rightarrow\infty$, the total number of qubits $N$
necessarily becomes infinite, but one has a considerable choice in how
to take this limit.  For example, we could use only one position bit,
thus fixing $n=1$, and let the number of momentum bits $N-1$ become
large. This is the limiting case of the Balazs-Voros-Saraceno
quantization. There is, however, a wide variety of other scenarios to
consider, e.g., $n=N/2$ or $n=2N/3-1$ as $N\rightarrow\infty$.  In
\cite{Tracy2002} it was shown that provided the number of momentum bits
$N-n$ approaches infinity, the correct classical behavior is recovered
in the limit.  If the number of momentum bits remains constant, i.e.,
$n=N-k$ ($k$ constant) as $N\rightarrow\infty$, a stochastic variant of
the classical baker's map is found. In the special case $n=N$ this
variant takes the form
\begin{equation}
s=\ldots s_{-2} s_{-1} s_0 \bullet s_1 s_2 s_3 \ldots \rightarrow
s^{\,\prime}=\ldots s_{-2} s_{-1} s_0\, r \bullet s_2 s_3 \ldots
\label{eq:BNNsymdyn}
\end{equation}
when written in the symbolic-dynamical language of Eq.~(\ref{eq:symbseq}).
The bit $r$ takes the value $s_1$ with probability
$\cos^2\left[\pi/2(0\cdot s_0s_{-1}s_{-2}\dots-1/2)\right]$ (and
$1-s_1$ otherwise). These results are consistent with those obtained
previously~\cite{Soklakov2000a}.

The extremal map has other interesting properties.  All
finite-dimensional unitary operators are quasi-periodic; the quantum
baker's map $\Bop_{N,N}$, however, is strictly periodic,
\begin{equation}
\left(\Bop_{N,N}\right)^{4N}=\one\;, \label{eq:B4N}
\end{equation}
as we show below.  All its eigenvalues, therefore, are $4N$-th roots of
unity, i.e., of the form $e^{\pi ik/2N}$, and hence, there are
degeneracies when $N>4$.  This represents a strong deviation from the
predictions of random matrix theory~\cite{Haake1991}. The eigenstates of
the extremal map were recently studied by Anantharaman and
Nonnenmacher~\cite{Anantharaman2005}, where $\Bop_{N,N}$ (with periodic
rather than antiperiodic boundary conditions) was called the
``Walsh-quantized'' baker's map.  The above degeneracy in the eigenvalues
allows constructions of eigenstates that remain partially localized in the
semiclassical limit, which means that ``quantum unique
ergodicity''~\cite{Rudnick1994} fails for this quantization.

The periodicity of the extremal map (\ref{eq:B4N}) can be easily shown after noting that $\Bop_{N,N}=-i\hat G_{N-1}\circ\hat S_N=
-i(\one_{2^{N-1}}\otimes\hat F_2)\circ\hat S_N$; i.e., $\Bop_{N,N}$ is
a shift followed by application of the unitary
\begin{equation}
\hat{C}\equiv-i\hat F_2=
{1\over\sqrt2}\Bigl(
e^{-\pi i/4}(\ket{0}\bra{0}+\ket{1}\bra{1})+
e^{\pi i/4}(\ket{0}\bra{1}+\ket{1}\bra{0})
\Big)
=e^{-\pi i/4}e^{(\pi i/4)\sx}\;, \label{eq:Cgate}
\end{equation}
which is a rotation by $90^\circ$ about the $x$ axis, to the least
significant position qubit. On product states, the action of
$\Bop_{N,N}$ can be written explicitly as
\begin{equation}
\Bop_{N,N}\ket{\psi_1}\otimes\ket{\psi_2}\otimes\dots\otimes\ket{\psi_N}=
\ket{\psi_2}\otimes\dots\otimes\ket{\psi_N}\otimes\hat{C}\ket{\psi_1}\;.
\label{eq:BN}
\end{equation}
Since $\hat C^4=1$, we get the property~(\ref{eq:B4N}). One can also see
that $\Bop_{N,N}$ cannot entangle initial product states.

When $n<N$, the action of the quantum baker's map is similar to
Eq.~(\ref{eq:BN}), but with a crucial difference.  After the qubit string
is cycled, instead of applying a unitary to the rightmost qubit, a
joint unitary is applied to all of the $N-n+1$ rightmost qubits.  As
discussed above, this joint unitary can be realized as controlled phase
change of the rightmost qubit, where the control is by the state
parameters $a_1,\ldots,a_{N-n}$ of the original momentum qubits.  This
controlled phase change means that initial product states become
entangled.

Entanglement production under iterations of the quantum baker's maps
was the subject of a recent paper~\cite{Scott2003}.  Since the
entangling controlled-phase change involves an increasing number of
qubits as $n$ decreases from $n=N$ to $n=1$ (the Balazs-Voros-Saraceno
map), one might expect that the entanglement increases as $n$ ranges
from $N$ to 1.  What was found, however, is that provided $n$ is not too
close to $N$, all the maps are efficient entanglement generators, but
the greatest entanglement is produced when $n$ is roughly midway
between $N$ and 1.  Starting with a uniform distribution of initial
product states, the mean entanglement ``quantum-baked'' into the
distribution was found to saturate at a level near to that
expected in random states. The small deviations from the entanglement
of random states might be due to hidden symmetries in the quantum baker's
maps~\cite{Abreu2006}.

Lastly, we mention another difference between the extremal quantum baker's
map $\Bop_{N,N}$ and other members of the baker's map family of
quantizations.  Ermann, Paz, and Saraceno~\cite{Ermann2006b} have found
that when a system with the dynamics of a quantum baker's map is cast in
the role of an environment acting on another quantum system, the extremal
quantum baker's map $\Bop_{N,N}$ is less effective at inducing decoherence
than other members of the family.  In particular, they showed that while
the entropy production rates of the different quantum baker's maps are
indistinguishable on a short time scale, which scales linearly with $N$,
$\Bop_{N,N}$ saturates much sooner than the other maps, thus displaying
the behavior expected for regular systems.

In view of the above described anomalous behavior experienced by the
extremal map, $\Bop_{N,N}$, our curiosity now invites an investigation
into the various currently prevailing perturbation-based tests for quantum
chaos, as applied to our class of quantum baker's maps. We start, however,
by investigating the simplest example, the extremal map $\Bop_{N,N}$
itself.

\section{Chaos in the extremal quantum baker's map?} \label{sec:extremalbakersmap}

\subsection{The extremal map and perturbations}
\label{sec:pert}

In Sec.~\ref{sec:bakersmap} we considered different quantizations of the baker's map
as unitary transformations, $\Bop_{N,n}$ ($n=1,\dots,N$), on a set of
$N$ qubits. When written in the form~(\ref{eq:Bshiftunit}), each of these
transformations consists of two steps: a cyclic shift, $\hat S_n$, in
which the $n$ leftmost qubits are shifted without otherwise being
altered and a unitary transformation on the rightmost $N-n+1$ qubits.
For the extremal quantum baker's map, $n=N$, which we consider in this
section, this second transformation is the gate $\hat{C}$ of
Eq.~(\ref{eq:Cgate}), which acts only on the single rightmost qubit as in
Eq.~(\ref{eq:BN}) and rotates it by $90^\circ$ about the $x$ axis.

We first examine the behavior of $\Bop_{N,N}$ under perturbations after
each iteration. In an effort not to affect qualitatively the dynamics of
the map itself, we choose our perturbations to be correlated across the
smallest possible distances in phase space. One choice might then be to
perturb only the single rightmost, least significant qubit in the position
basis. Indeed, such a choice leads to the smallest changes in position. As
a consequence of the uncertainty principle, however, perturbing the {\it
least\/} significant qubit in the position basis causes correlated changes
across the {\it greatest\/} distances in momentum.  In opposition to our
classical intuition, no single qubit can be thought of as being ``more
significant'' than another in an overall phase-space sense.  Our
particular choice of qubit upon which to perturb does not affect the
phase-space area of correlated changes made to a state; however,
perturbations affecting the middle qubit(s) give rise to correlated
changes across the smallest phase-space distances. In the present
analytical study, it is simplest to take the rightmost qubits as being the
least significant.  These considerations are revisited later in our
numerical investigations, where we instead choose to perturb the middle
qubit(s).

Suppose that with each step we perturb the $m$ rightmost qubits, where
$m\ll N$, by applying an $m$-qubit unitary transformation $\hat U_k^{(m)}$
chosen at random. For the moment, we assume that {\it any\/}
transformation is allowed, but the arguments still work even if only a
finite set of transformations is allowed.

For simplicity, we assume that the system is initially in the
tensor-product state $\ket0^{\otimes N}$.  Suppose that $m=1$ and that
the perturbation affects only the single rightmost qubit.  Then the
perturbation operators are all of the form $\one_{2^{N-1}}\otimes
\Uop_k^{(1)}$, and the state after a single step becomes
\begin{equation}
\ket0^{\otimes N} \rightarrow
  \ket0^{\otimes N-1} \otimes \left( \Uop_{k_1}^{(1)}\Cop\ket0 \right) \;,
\end{equation}
while the next step transforms it to
\begin{equation}
\ket0^{\otimes N-1} \otimes  \left( \Uop_{k_1}^{(1)}\Cop\ket0 \right) \rightarrow
  \ket0^{\otimes N-2} \otimes \left( \Uop_{k_1}^{(1)}\Cop\ket0 \right)
  \otimes \left( \Uop_{k_2}^{(1)}\Cop\ket0 \right) \;,
\end{equation}
and so forth.  It is clear that the above dynamics does not explore the
entire Hilbert space, since the state remains a tensor product as long as the
perturbation is restricted to a single qubit.

The situation changes if we let $m=2$.  Since two-qubit gates between
nearest neighbors are sufficient for universal quantum computation, any
state can be produced by the shift map plus two-qubit perturbations. It
does not follow, however, that all states can be reached {\it quickly\/};
in general, the number of gates needed to reach a generic state of $N$
qubits increases exponentially with $N$, which implies that many
``rounds'' (complete sets of $N$ steps) are needed to reach most states.
On the other hand, as we saw in Sec.~\ref{subsubsec:quanthyper}, the
perturbation need not sample generic vectors to elicit evidence for
hypersensitivity, so considerations of universality in quantum computation
and the time needed to sample generic states provide little information
about hypersensitivity.

\subsection{Signatures of chaos for $\Bop_{N,N}$}   \label{sec:shiftSignatures}

In this section we show that the extremal quantum baker's map is chaotic
according to two popular signatures of quantum chaos: it displays a linear
entropy increase when coupled to an environment, and the fidelity between
two vectors evolving according to the original map and a slightly changed
version of the map decreases exponentially.  It does not, however, display
hypersensitivity to perturbation.

\subsubsection{Fidelity decay}
\label{subsec:fidelity}

Define a modified baker's map by
\begin{equation}
\Bop_{N,N}^{\,\prime} \ket{x_1}\otimes\cdots\otimes\ket{x_N} \equiv
\ket{x_2}\otimes\cdots\otimes\ket{x_N}\otimes \Uop^{(1)}\Cop\ket{x_1} \;,
\end{equation}
where $\Uop^{(1)}$ is a single-qubit unitary map satisfying
$0<|\bra{0}\Cdag\Uop^{(1)}\Cop\ket{0}| < 1$.   We can define $\lambda = -2\ln
|\bra{0}\Cdag\Uop^{(1)}\Cop\ket{0}|>0$.  Letting
$\ket{\psi(t)}=\big(\Bop_{N,N}\big)^t\ket{\psi_0}$ and
$\ket{\psi^\prime(t)}=\big(\Bop_{N,N}^{\,\prime}\big)^t\ket{\psi_0}$,
where $\ket{\psi_0}=\ket0^{\otimes N}$, we see that
the fidelity decreases exponentially with the number of iterations:
\begin{equation}
F(t)=|\bra{\psi^\prime(t)}\psi(t)\rangle|^2 = e^{-\lambda t} \;,
\quad t=0,\ldots,N.
\end{equation}
For greater numbers of iterations, the simple exponential decay is
modified as the qubits experience more than one application of $\Cop$
or $\Uop^{(1)}\Cop$.

\subsubsection{Linear increase of entropy}
\label{subsec:entropyincrease}

Let the environment be a collection of qubits in the maximally mixed
state.  After each iteration of the map, the register interacts with a
fresh environment qubit.  The interaction is given by a controlled $\sx$
operation, with the environment qubit acting as control and the target
being the rightmost system qubit. In the notation of
Sec.~\ref{sec:hyperdef}, this binary perturbation amounts to an
application of one of $\hat{U}_0=\one_{2^N}$ or
$\hat{U}_1=\one_{2^{N-1}}\otimes\sx$ (chosen with equal probability) at
each time step (this stochastic perturbation is the $m=1$ model of
Sec.~\ref{sec:pert}, with the single-qubit perturbation unitaries restricted to
the identity and $\sx$). After tracing out the environment, one iteration
of the perturbed map is described by the quantum operation
\begin{equation}
{\cB}(\r) = {1\over2} \Bop_{N,N}\r \Bdag_{N,N} +
{1\over2} \left(\one_{2^{N-1}}\otimes\sx\right)\Bop_{N,N}\r
  \Bdag_{N,N}\left(\one_{2^{N-1}}\otimes \sx\right) \;.
\end{equation}
For the initial state $\r_0=(\ket0\bra0)^{\otimes N}$, and denoting by
${\cB}^t$ the $t$-th iterate of ${\cB}$, we have
\begin{equation}
{\cB}^t(\r_0) = (\ket0\bra0)^{\otimes(N-t)}\otimes
                      \left(\one_2/2\right)^{\otimes t} \;,
\end{equation}
since $\sx$ commutes with $\hat C$.
The entropy of ${\cB}^t(\r_0)$ is $t$ bits.  The entropy thus increases
at a rate of 1 bit per iteration until it saturates at $N$ bits after
$N$ steps.  Under single-qubit perturbations, it is clear that the perturbed
vectors explore an exponentially increasing number of
Hilbert-space dimensions, but it is equally clear that they do not
explore these dimensions randomly.

\subsubsection{Hypersensitivity to perturbation}

{From} the discussions in Secs.~\ref{subsubsec:quanthyper} and
\ref{sec:pert}, it can be seen that $\Bop_{N,N}$ is not hypersensitive to
perturbations that affect only the single rightmost qubit. For an extreme
example of this, consider the binary perturbation in
Sec.~\ref{subsec:entropyincrease} immediately above.  After $t\le N$ map
iterations, there are $2^t$ perturbation histories, which, with the
initial state $\ket{\psi_0}=\ket{0}^{\otimes N}$, correspond to orthogonal
system vectors:
\begin{equation}
\ket{\psi_\bfk}=\ket{0}^{\otimes (N-t)}
\hat{C}\ket{k_1}\otimes\cdots\otimes\hat{C}\ket{k_t}\;,
\end{equation}
where $k_i\in\{0,1\}$, in the notation of Sec.~\ref{sec:hyperdef}.  A
measurement on the environment that groups these vectors according to the
values of $k_1,\dots,k_j$, with $0\le j\le t$, reduces the average system
entropy from $H_{\cS}=t$ to $H(\Imin=j)=t-\Imin$ bits.  When $t=N$, the perturbed
system vectors make up an orthonormal basis, and for $t\ge N$, the
perturbation produces $2^{t-N}$ copies of an orthonormal basis.  Thus, for
$t\ge N$, the information-entropy trade-off relation is
$H(\Imin)=N-\Imin$.  For all $t$, our hypersensitivity parameter takes the
value $s=1$.  In this extreme example, each bit of information purchases a
bit of entropy reduction, as is always true when the perturbed vectors are
drawn from an orthonormal basis with each vector in the basis having the
same overall probability.

In general, stochastic perturbations that affect only a single qubit of
the extremal quantum baker's map are expected to produce an
information-entropy trade-off that is linear (to good approximation) near
the maximal entropy.  Although the hypersensitivity parameter generally
varies with both the choice of perturbation and number of map iterations,
its magnitude should not exceed 5.3, the bound on~$\rs$ for product states
[Eq.~(\ref{eq:s2})]. The extremal map $\Bop_{N,N}$, therefore, does not
exhibit exponential hypersensitivity to perturbation under single-qubit
perturbations.  By contrast, we have seen above that it does exhibit
linear growth of entropy and exponential decay of fidelity.
Hypersensitivity to perturbation is evidently a finer sieve than the other
two perturbation-based criteria.

The reason the perturbed extremal map does not explore Hilbert space
efficiently is that the map itself produces no entanglement. In contrast,
the nontrivial quantizations of the baker's map are efficient entanglement
generators~\cite{Scott2003}, producing entanglement that saturates after
several iterations at a level close to that expected in random states. For
these nontrivial quantizations, even a single-qubit perturbation, together
with the entangling transformation of the unperturbed map, generically
gives rise to a universal set of unitary gates, so in time the system can
approach any state in the Hilbert space.  Although the speed at which this
happens remains unknown, our numerical results for hypersensitivity to
perturbation, presented in the next section, suggest that if $n$ is not
too close to $N$, the perturbed nontrivial quantizations do efficiently
explore all of Hilbert space.

Both the simple analytical argument above and the numerical results in the
next section are for single-qubit perturbations. A systematic study of
hypersensitivity to perturbations acting on two or more qubits is beyond
our current numerical capabilities. In the remainder of this section, we
present an analytical argument that suggests that, for a small number of
time steps and for maps close to the extremal map $\Bop_{N,N}$, the
information-entropy trade-off is linearly bounded even for entangling
perturbations acting on two qubits.

We choose a perturbation that affects the two rightmost qubits, i.e.,
$k=2$. Given some reasonable assumptions about the stochastic
perturbation, if we average over all perturbations, the state after $t$
steps is approximately equal to
\begin{equation} \r^{(N)} \approx \left(
  \outer00 \right)^{\otimes N-t-1}
  \otimes \left( \one_2/2 \right)^{\otimes t+1} \;. \label{eq:densop}
\end{equation}
This state has von Neumann entropy $H_{\cS}=t+1$.  We would like to
acquire some information $\Imin$ about the perturbations which enables us
to reduce this entropy by a small amount $\Ds= H_{\cS}-H$.

We now show that the ratio $\Imin/\Ds$ is bounded above by a quantity
that is independent of $t$ for all $t<N$. Suppose that after $t$ steps,
our system is in state (\ref{eq:densop}). Now let us apply $\Bop_{N,N}$,
but {\it not\/} the perturbation.  If we trace out all but the two least
significant qubits, these two qubits are in the state \begin{equation}
\r^{(2)} = \one_2/2 \otimes \Cop\outer00\Cdag \;, \end{equation} which has
one bit of entropy.  The perturbation affects only these two bits, so the
state of the other $N-2$ qubits is irrelevant to the entropy increase.
Now we apply the perturbation and get \begin{equation} \r^{(2)}
\rightarrow \sum_\xi p_\xi \Uop_\xi \r^{(2)} \Udag_\xi
  = \left( \one_2/2 \right)^{\otimes 2} \;,
\end{equation}
where $\xi$ labels which perturbation is performed, $p_\xi$ is the
probability of that perturbation, and $\Uop_\xi$ is the corresponding
two-qubit unitary transformation.  The entropy of the new state is two
bits, giving an entropy increase of one bit.

Clearly, we can reduce the entropy by one bit if we can determine which
$\Uop_\xi$ was actually performed.  If the perturbations are drawn from a
discrete set, the number of bits needed to determine this is given by the
entropy of the distribution $p_\xi$, i.e., \begin{equation} \Imin \le -
\sum_\xi p_\xi \log p_\xi \;. \end{equation} If $\xi$ is continuous, then
{\it fully\/} determining $\Uop_\xi$ would require an infinite amount of
information.  Since the space of two-qubit operators is not very large,
however, it doesn't take that much information to know $\Uop_\xi$ to a
good approximation; e.g., we could achieve an entropy reduction of nearly
a bit at a cost of approximately 45 bits by knowing each of the 15
relevant parameters of an arbitrary two-qubit unitary with three bits of
precision.

This procedure, while not necessarily optimal, places a rather low bound
on the ratio $\Imin/\Ds$, a bound independent both of the number of
iterations, $t$, and the number of qubits, $N$.  This argument changes
little if we use $\Bop_{N,N-1}$ instead of $\Bop_{N,N}$, or $\Bop_{N,N-k}$
for $k$ small compared to $N$.  Nor does it change much if the
perturbation affects $k$ bits, so long as $k$ is small compared to $N$.
If the perturbation affects {\it many\/} bits, however, or if a
quantization $\Bop_{N,N-k}$ is used for {\it large\/} $k$, the upper bound
on $\Imin/\Ds$ becomes so large that it gives little restriction.

The above provides some evidence for the conjecture that maps close to the
extremal map, $\Bop_{N,N}$, do not exhibit exponential hypersensitivity to
entangling perturbations. Since these results are valid only as long as
$t\le N$, i.e., as long as the number of iterations does not exceed the
number of qubits, this evidence must be regarded as suggestive, but
inconclusive.

\section{Numerical results}    \label{sec:numerical}

We now investigate numerically the entire class of quantum baker's maps,
$\hat B_{N,n}$ ($n=1,\dots,N)$, in the context of the three
perturbation-based criteria for quantum chaos. As remarked in
Sec.~\ref{sec:pert}, perturbations affecting the middle qubit(s) cause
correlated changes to the state in phase space across the smallest
possible distances.  Up until now we have applied all perturbations to the
rightmost, least significant qubit in position. Since the application of
$\hat C^{\dag}$ as a perturbation to this qubit would undo the dynamics in
momentum in the case of the extremal quantum baker's map, $\Bop_{N,N}$,
one might judge this perturbation to be atypical, upsetting the crucial
momentum dynamics of the map.  To avoid this, we choose henceforth the
total number of qubits $N$ to be odd, and we perturb the middle qubit.

The perturbation we choose for this qubit is a simple binary
perturbation, a rotation by angle $\pm2\pi\alpha$ about the $y$ axis,
\begin{equation}
\hat U_k(\alpha)
\equiv
\one_{2^{(N-1)/2}}\otimes
e^{\pi i(-1)^k\alpha\sy}
\otimes\one_{2^{(N-1)/2}}\;.
\label{eq:pert}
\end{equation}
The perturbation is conditioned on the binary
environmental states $\ket{k}_E$, $k=0$ or 1. To be precise, after each
iteration of the map, the system couples to its environment through a
joint conditional evolution with end result
\begin{equation}
\r_{\text{total}} =\frac{1}{2} \left[\left(\hat U_0 \r\hat U_0^\dag\right) \otimes\ket{0}_E\bra{0}+\left(\hat U_1 \r \hat U_1^\dag\right) \otimes\ket{1}_E\bra{1}\right].
\end{equation}
To avoid inapt comparisons, we use  $\alpha=0.2$ (rotation angle $0.4\pi$)
and initial system state $\ket{\psi_0}=\ket{0}^{\otimes N}$ throughout
this section.

\begin{figure}[t]
\includegraphics[width=14cm]{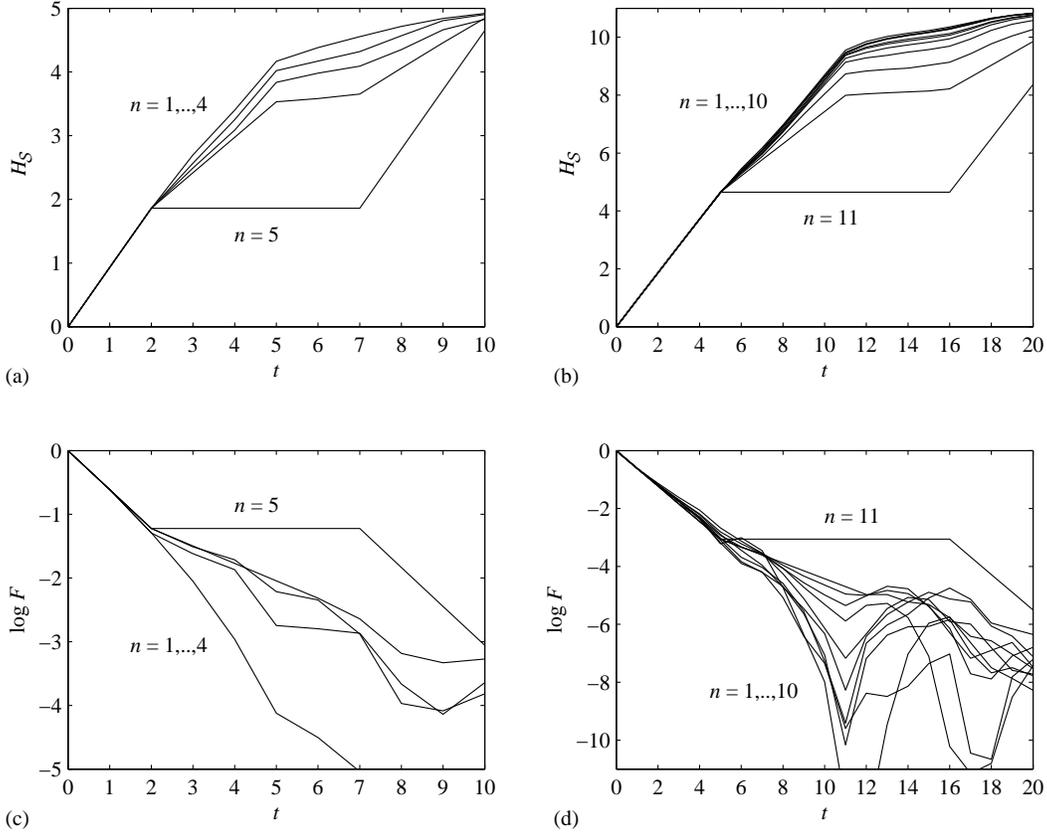}
\caption{The rate of increase in entropy is initially linear for
all (a) five-qubit and (b) eleven-qubit quantizations of the baker's map. The rate of
decrease in fidelity is initially exponential for all (c) five-qubit and (d) eleven-qubit
quantizations.}
\label{fig2}
\end{figure}

We can cope with the unwanted perturbation in many different ways. One
possibility is to accept an increase in entropy and average over all
perturbation histories by tracing out the environment. The system entropy,
$H_{\cS}=-\tr(\r\log\r)$, then increases at an initially constant linear
rate for all quantum baker's maps. This is shown for the quantizations
using $N=5$ qubits in Fig.~\ref{fig2}(a) and $N=11$ qubits in
Fig.~\ref{fig2}(b). The rate of entropy production for the different
quantizations is nearly the same for the first $t=(N-1)/2$ iterations. In
contrast to the other quantizations, the entropy produced by the extremal
map, $\Bop_{N,N}$, remains constant for times beyond $t=(N-1)/2$ before
resuming its climb towards the maximal entropy of 5 or 11 bits. Thus,
although there is a quantitative change in entropy production at later
times, the different baker's maps behave qualitatively the same. These
numerical results support the simple analysis of
Sec.~\ref{subsec:entropyincrease}.

Alternatively, if the above entropy production proves unacceptable, we
could instead perform a measurement on the environment at each time
step to record which perturbation actually occurs.  Consider the
perturbation history of all 1's.  The fidelity decay
between two initially equal quantum states that evolve either according to
this extremal perturbation history or the unperturbed map,
\begin{equation}
F(t)=\Bigl|\bigl\langle\psi_0\bigl|\Bigl(\Bdag_{N,n}\hat U_1^\dag\Bigr)^t
    \Bigl(\Bop_{N,n}\Bigr)^t\bigr|\psi_0\bigr\rangle\Bigr|^2\;,
\end{equation}
might be used as an indicator of the underlying dynamics of the map. The
rate of fidelity decay for all quantum baker's maps is initially
exponential. This is shown in Figs.~\ref{fig2}(c) and \ref{fig2}(d) for
the five-qubit and eleven-qubit quantizations. Although the fidelity
corresponding to the extremal map, $\Bop_{N,N}$, stalls at approximately
$t=(N+1)/2$ iterations, all quantizations are found to exhibit decay rates
which are initially exponential. Again, these numerical results support
the simple analysis of Sec.~\ref{subsec:fidelity}.

The iteration at which the entropies and fidelity decays first become
appreciably different for the various quantizations remains at $t=(N+1)/2$
as $N$ increases, and thus our conclusions become stronger in the limit of
large $N$.  To keep our analysis strictly in the quantum regime, however,
we focus on the five-qubit quantizations for the remainder of this
section.

To investigate hypersensitivity to perturbation, we first consider a
particularly intuitive algorithm for grouping vectors, which is based on
finding structure produced by the temporal order of the perturbations.
Each grouping corresponds to measuring, after a fixed number of iterations
$t$, the environment states---and, hence, the applied perturbation---at
$l\le t$ times.  The $2^t$ perturbation histories---and their final
states---are thus grouped into $2^l$ sets, each containing $2^{t-l}$
states.  It takes $\Ibar=l\;$bits to specify a group.

As an example of this procedure, suppose we have $t=4$ iterations and
we choose to measure the first and last states of the environment.
Thus $l=2$, and all histories are grouped into $2^l=4$ sets of
$2^{t-l}=4$ binary strings in the form $0**\:0$, $1**\:0$, $0**\,1$ and
$1**\,1$, where $*$ denotes an arbitrary entry.  Defining
\begin{equation}
\ket{k_1k_2\cdots k_t}\equiv \hat U_{k_1}\Bop_{N,n}\hat U_{k_2}
         \Bop_{N,n}\cdots\hat U_{k_t}\Bop_{N,n}\ket{\psi_0}\;,
\end{equation}
the final state of the system, conditioned on measurement results $i$
and $j$ for the first and last environment qubits, is
\begin{equation}
\r_{ij}=2^{l-t}\sum_{k_2,k_3\in\{0,1\}}\ket{i k_2k_3 j}
    \bra{i k_2k_3 j} \;.
\label{eq:timegrouping}
\end{equation}
Consequently, at the expense of storing $l=2$ bits of information, we
can, on average, reduce the entropy to
\begin{equation}
\Hbar = -{1\over2^l}\sum_{i,j\in\{0,1\}}\tr(\r_{ij}\log\r_{ij})\;.
\end{equation}

The particular two bits stored in this example might not be the optimal
choices. There are countless other measurements to consider, some of which
no doubt lead to lower average entropies. For the moment, however, we
restrict our measurements to the above type and minimize $\Hbar$ over the
$\,t\,\choose l$ possible choices for the measurement times.  Denoting
this minimum entropy by $H$, the minimum information needed to reduce the
average system entropy to $H$ is then $\Imin = l$ bits. Although there is
no guarantee that $l$ is in fact the overall minimum, this simple scheme,
which we call the \emph{temporal\/} grouping algorithm, proved superior to
previously used schemes (e.g., those discussed in~\cite{Schack1996b}) for
the maps and perturbations considered here.

\begin{figure}[t]
  \includegraphics[width=14cm]{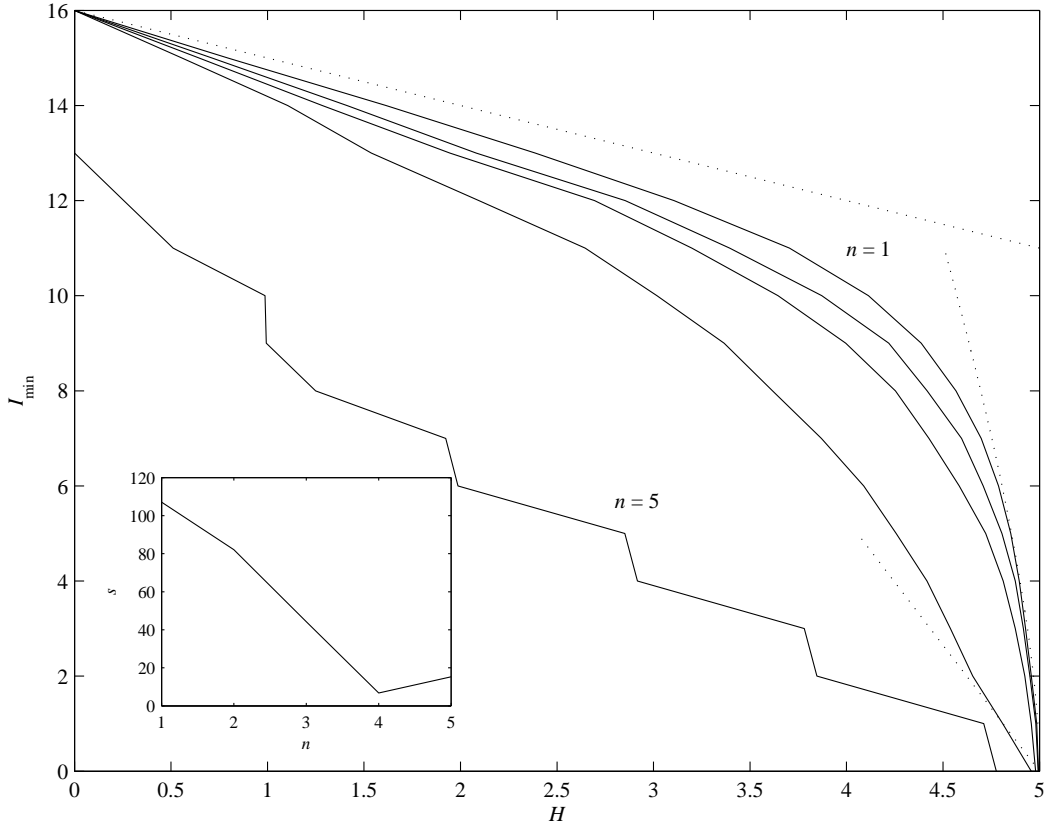}
\caption{The minimum information $\Imin$ needed to
  reduce the entropy to ${H}$
  after 16 map iterations, using the temporal grouping algorithm, for all perturbed five-qubit quantizations of the baker's map ($n=1,\ldots,5$).
  The upper and rightmost dotted curves are the approximate sphere-grouping trade-off for random vectors
  in $d=32$ dimensions [Eq.~(\ref{eq:tradeoff})], while the lower dotted line is the linear trade-off for random product vectors
  [Eq.~(\ref{eq:tradeoff2})].  The inset shows the hypersensitivity parameter~$\rs$ for each quantization.}
\label{fig3}
\end{figure}

\begin{figure}[t]
  \includegraphics[width=14cm]{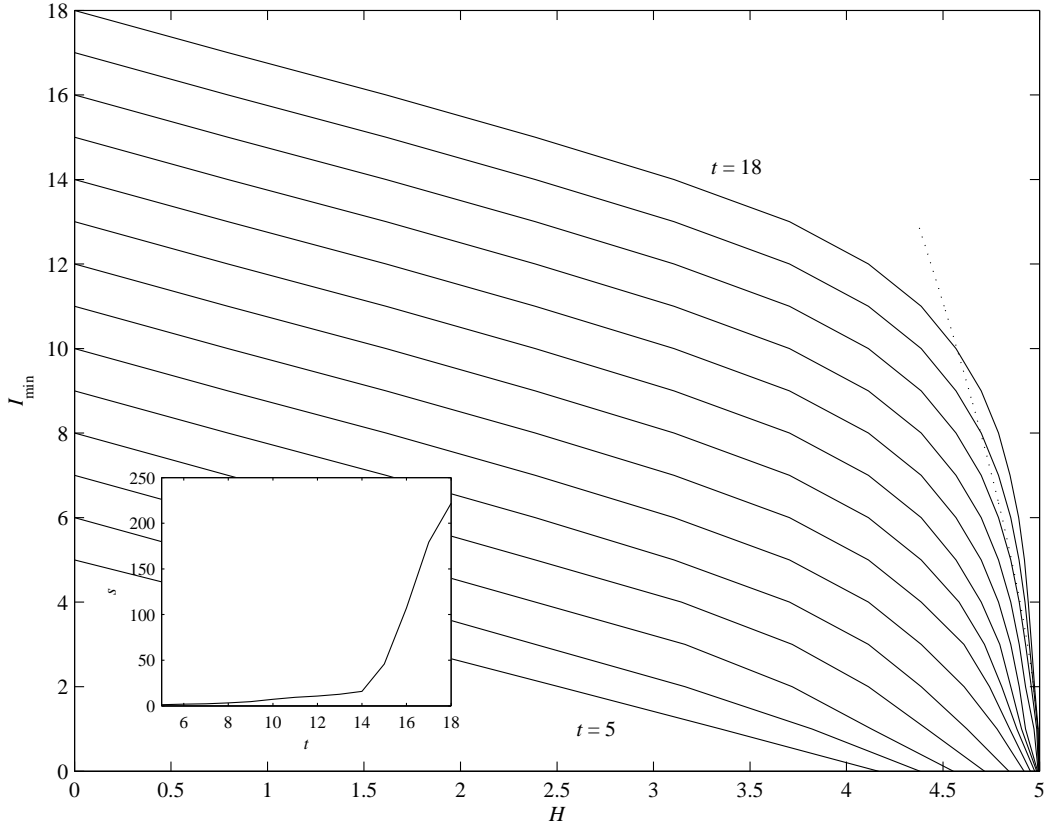}
\caption{The minimum information $\Imin$ needed to
  reduce the entropy to ${H}$ after $t$ map iterations, where $5\le t\le18$, for the perturbed five-qubit
  Balazs-Voros-Saraceno quantization of the baker's map ($n=1$), using the temporal grouping algorithm.
  The dotted curve is the sphere-grouping trade-off for random vectors in 32 dimensions. The inset shows
  the hypersensitivity parameter~$\rs$ at each iteration.}
\label{fig4}
\end{figure}

\begin{figure}[t]
  \includegraphics[width=14cm]{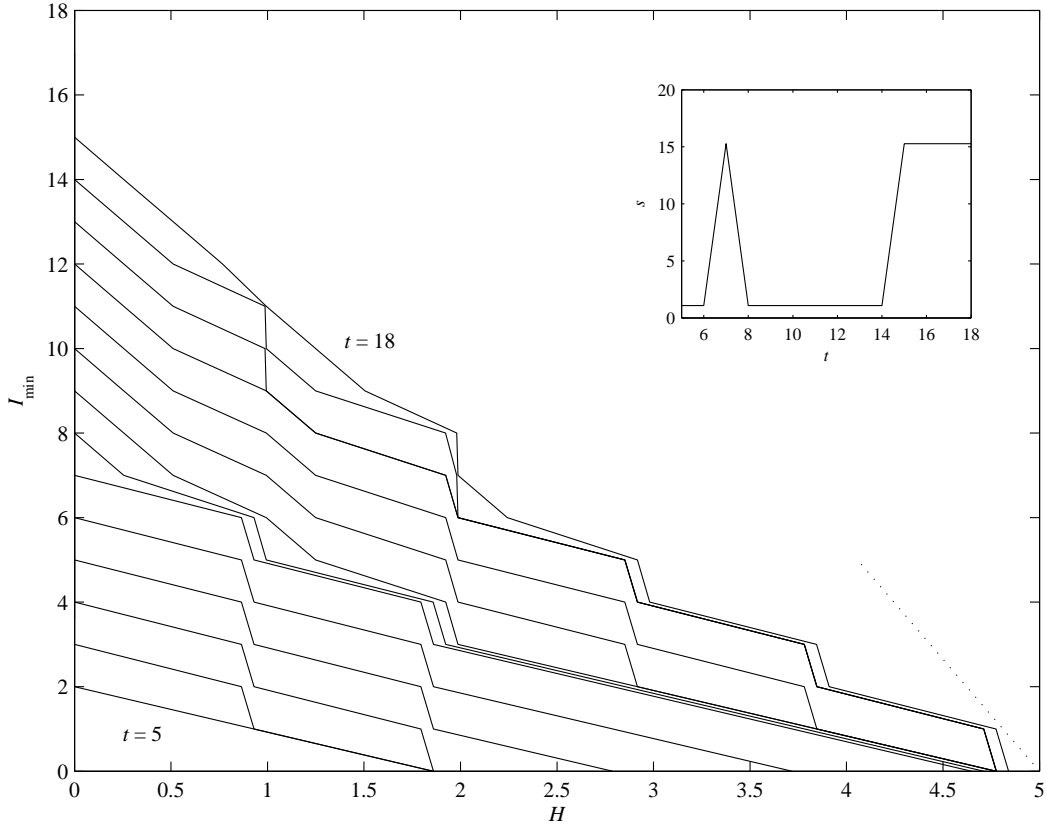}
\caption{The minimum information $\Imin$ needed to
  reduce the entropy to ${H}$ after $t$ map iterations, where $5\le t\le18$, for the perturbed five-qubit
  extremal quantization of the baker's map ($n=N$), using the temporal grouping algorithm. The dotted
  line is the trade-off for random product vectors [Eq.~(\ref{eq:tradeoff2})]. The inset shows the
  hypersensitivity parameter~$\rs$ at each iteration.}
\label{fig5}
\end{figure}

\begin{figure}[t]
  \includegraphics[width=14cm]{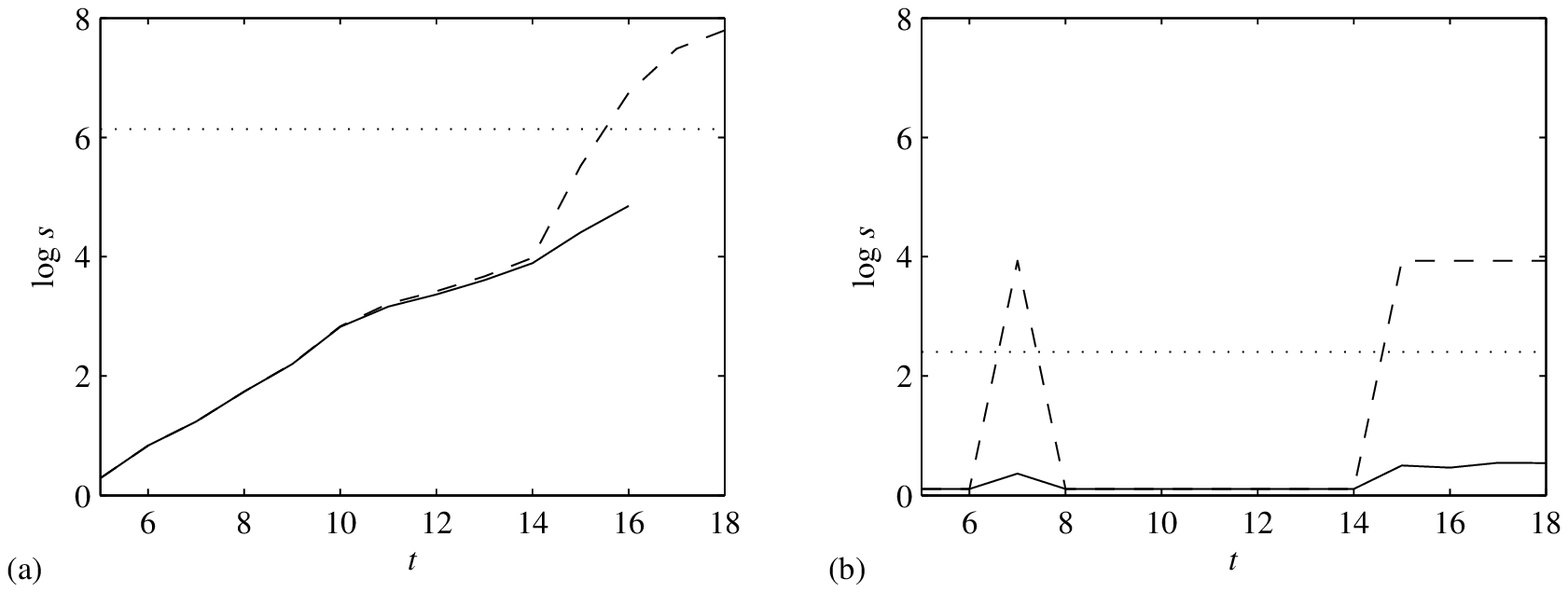}
\caption{The hypersensitivity parameter~$\rs$ after $t$ map iterations, where $5\le t\le18$,
for the perturbed five-qubit (a) Balazs-Voros-Saraceno ($n=1$) and (b)
extremal ($n=N$) quantizations of the baker's map. The spurious higher values of~$\rs$
arising from the temporal groupings (dashed lines) are significantly reduced
using a genetic-algorithm approach (solid lines). The dotted lines are the approximate upper
bounds on~$\rs$ corresponding to (a) random vectors [Eq.~(\ref{eq:s})] and (b) random product
vectors [Eq.~(\ref{eq:s2})].
}
\label{fig6}
\end{figure}

Using this procedure, in Fig.~\ref{fig3} we plot $\Imin$ versus $H$ for
all perturbed five-qubit quantum baker's maps after $t=16$ map iterations
(solid lines). The perturbing parameter remains at $\alpha=0.2$ and initial
state at $\ket{\psi_0}=\ket{0}^{\otimes N}$. The region of interest
regarding the question of quantum chaos lies to the right where $\Imin$ is
small. Here we see that, except for the quantizations with $n$ close to
$N=5$, a very large amount of information is required to reduce the system
entropy by a small amount. This is a distinguishing characteristic of
chaos, which is absent for the extremal quantization ($n=N$).  When
$\Imin$ is small, the information-entropy trade-off is characterized by
the hypersensitivity parameter~$\rs$ [Eq.~(\ref{eq:iquantity})]. The inset
in Fig.~\ref{fig3} shows this quantity for all five quantizations. Recall
that $1/\rs$ is the reduction in system entropy purchased by gathering one
bit of information about the environment. Although entropy reduction is
affordable for the extremal quantization, one bit buys very little when
$n$ approaches 1. The dotted lines show our theoretical trade-offs for
random vectors (upper and rightmost) and random product vectors (lower),
given by Eqs.~(\ref{eq:tradeoff}) and (\ref{eq:tradeoff2}), respectively.
When $n$ approaches 1, the information-entropy trade-off approaches that
expected for random vectors, while for $n=N$, it is bounded by the
trade-off for random product vectors.

Using the same grouping algorithm, we plot in Fig.~\ref{fig4} the
information-entropy trade-off for a growing number of iterations of the
Balazs-Voros-Saraceno quantization ($n=1$). The figure shows $\Imin$
versus $H$ for 5--18 iterations of $\Bop_{5,1}$, and in the inset, the
corresponding value of~$\rs$. To a rough approximation, our
hypersensitivity signature~$\rs$ appears to grow exponentially with the
number of iterations. This map thus exhibits numerical evidence of
exponential hypersensitivity to perturbation. Notice, however, that the
trade-off violates the sphere-grouping bound (\ref{eq:tradeoff}) derived
from random states for $t\gtrsim 16$ (dotted line), which means the
current method for gathering information about the environment is not
optimal, a situation we discuss further below.  The extremal quantization
displays a strikingly different behavior. The information-entropy
trade-off and the parameter~$\rs$ for 5--18 iterations of $\Bop_{5,5}$ are
shown in Fig.~\ref{fig5}. In this case the information-entropy trade-off
remains approximately linear for all levels of iteration, with a very
roughly constant~$\rs$.  There is no evidence of hypersensitivity to
perturbation.

We now investigate the hypersensitivity parameter in greater detail for
the Balazs-Voros-Saraceno and extremal quantizations. The graphs of~$\rs$
in the insets of Figs.~\ref{fig4} and \ref{fig5} are redrawn, now on a
logarithmic scale, as the dashed lines in Figs.~\ref{fig6}(a) and
\ref{fig6}(b), respectively. The horizontal dotted lines in each of these
figures are the upper bounds~(\ref{eq:s}) and (\ref{eq:s2}), respectively,
corresponding to the values of~$\rs$ for random vectors and random product
vectors. Notice that in both cases the dashed lines cross these bounds.
This indicates that the temporal groupings used up until now are not
optimal. Indeed, in the case of the Balazs-Voros-Saraceno quantization, by
considering groupings which correspond to partitions of projective Hilbert
space into two equal volumes, we find that one bit of information can buy
larger entropy reductions when $t\geq 16$.  This method of grouping vectors,
however, works well only for distributions that are close to random.

We now consider grouping algorithms that are not constrained by a supposed
temporal structure of the vector distribution. Although optimal groupings
can always be found by simply testing every possibility, the size of the
search space is doubly exponential in the number of map iterations. We
thus turn to the theory of combinatorial optimization. Specifically, a
simple genetic algorithm~\cite{Eiben2003} was used to partition vectors
into two groups with the goal of minimizing the average conditional
entropy $\Hbar$. Although these groups were not constrained to be of equal
size, the returned solution always corresponded to $\Ibar=1\pm 0.003$, and
thus we can take $H(\Imin=1)=\Hbar$ to a very good approximation. The
corresponding value of~$\rs$ for this method is plotted as the solid line
in Fig.~\ref{fig6}. In many cases the genetic algorithm located precisely
the same vector grouping that was found previously by the temporal
grouping algorithm. The spurious higher values of~$\rs$, however, are now
significantly reduced for both quantizations. Although Fig.~\ref{fig6}(a)
remains incomplete due to computational constraints, for the data points
calculated, $\log\rs$ has regained its linear approach to the upper bound,
where it eventually will saturate.  The difference between the two
quantizations under single-qubit perturbations is now difficult to
dispute.  The criterion of hypersensitivity to perturbation thus
unmistakeably distinguishes the dynamics of the extremal quantum baker's
map ($n=N$) as qualitatively different from the Balazs-Voros-Saraceno
quantization ($n=1$).

\section{Conclusion}
\label{sec:conclusion}

This paper addresses the difficult question of how to characterize quantum
chaos dynamically in the hard quantum regime, far from the classical limit
where signatures of the classical sensitivity to initial conditions can be
identified in the quantum properties of a system.  In this hard quantum
regime, criteria for quantum chaos rely on studying the effects of
perturbing the quantum dynamics.  In this paper we study the three
perturbation-based criteria that have been proposed: linear increase of
entropy when a system is coupled to a perturbing environment; exponential
decay of the fidelity between the unperturbed dynamics and a modified
dynamics; and hypersensitivity to perturbation under stochastic
perturbations of the dynamics.  Hypersensitivity to perturbation is
formulated in terms of the entropy reduction achieved by acquiring
information about the perturbation, which we call the information-entropy
trade-off.  Of these three criteria, hypersensitivity to perturbation is
by far the most difficult to define rigorously and to investigate
analytically and numerically.

We apply these three criteria to a set of qubit-based quantizations of the
baker's map.  These quantizations range from a map that has a trivial
unentangling shift dynamics to the original Balazs-Voros-Saraceno
quantized map, which is highly entangling.  We find, through a combination
of analytical arguments and numerical results, that all the quantizations
exhibit a linear entropy increase and an exponential fidelity decay.  In
contrast, we show that the criterion of hypersensitivity to perturbation
distinguishes the entangling quantizations from the shift map.  In
particular, by focusing on the trivial shift map and the
Balazs-Voros-Saraceno map, our numerical work on hypersensitivity provides
compelling evidence that these maps behave quite differently under
stochastic perturbations, as revealed by studying the information-entropy
trade-off for these maps.

The reason that hypersensitivity to perturbation is different from the
other two perturbation-based criteria is not hard to identify.  Linear
entropy increase and exponential fidelity decay both tell one about how
widely a perturbation disperses vectors in Hilbert space, but they provide
no information about how randomly the perturbed vectors populate Hilbert
space.  In contrast, the randomness of the distribution of perturbed
vectors is precisely what the information-entropy trade-off is sensitive
to.

One way to quantify the trade-off in a single number is provided by the
hypersensitivity parameter~$\rs$: $\rs^{-1}$ is {\it defined\/} in terms of
the information-entropy trade-off as the entropy reduction purchased by an
optimal one bit of information about the perturbation, but $\rs$ can be
{\it interpreted\/} as the number of Hilbert-space dimensions explored
randomly by the perturbed vectors.  Thus when~$\rs$ increases exponentially,
as our numerical work indicates for the Balazs-Voros-Saraceno map, it
signals that the perturbed vectors are populating an exponentially
increasing number of dimensions in a random way.  Linear entropy increase
and exponential fidelity decay do not provide information about this
property of chaotic quantum dynamics.

The numerical hypersensitivity results in this paper are obtained for
single-qubit perturbations of the five-qubit baker's maps.  Investigating
more general perturbations would involve dealing with more qubits and thus
would require considerably greater computational resources.  To reduce the
required computational resources in future investigations of
hypersensitivity, it would be highly desirable to have an analytical
argument or sufficient numerical evidence to demonstrate convincingly that
the hypersensitivity parameter~$\rs$ is, by itself, a reliable signature
of hypersensitivity to perturbation.  Were this established, numerical
investigations of hypersensitivity could be reduced from computing the
entire information-entropy trade-off to calculating the trade-off only for
the case of one bit of acquired information.

\begin{acknowledgments}
This work was supported in part by U.S.~Office of Naval Research Grant
No.~N00014-00-1-0578, U.S.~National Science Foundation Grant
No.~CCF-0448658, and the European Union IST-FET project EDIQIP.  AJS
acknowledges support from CIAR, CSE, iCORE and MITACS.
\end{acknowledgments}

\appendix
\section{Information-entropy trade-off for random vectors}
\label{app:tradeoff}

Consider $\cN$ state vectors distributed randomly in a
$\ccD$-dimensional Hilbert space, where we assume that $\cN\ge\ccD$.
Given an entropy $H\le\log\ccD$, we group the vectors into groups
that on average have this entropy and then ask how much
information is required to specify a group.  For a given $H$, we
are interested in the grouping that minimizes the required
information, as in Eq.~(\ref{eq:defimin}).  The relation between
$\Imin$ and $H$ is the {\it information-entropy trade-off}.  In
this Appendix we formulate an approximate information-entropy
trade-off for random vectors by grouping the vectors into spheres
on projective Hilbert space whose radius is given by a
Hilbert-space angle $\phi$.  This being an approximate trade-off
relation, we denote the information by $I$ instead of $\Imin$.

The sphere-grouping model is based on results, given
in~\cite{Schack1994a}, for the volume and entropy of a Hilbert-space
sphere.  The model was formulated in~\cite{Schack1996b} and refined
in~\cite{Soklakov2000b}.

The number of spheres of radius $\phi$ that can be packed into
projective Hilbert space is given by Eq.~(A18) of~\cite{Schack1994a},
\begin{equation}
\ND={\cV_{\ccD}\over\cV_{\ccD}(\phi)}=(\sin^2\!\phi)^{-(\ccD-1)}\;,
\end{equation}
where $\cV_{\ccD}(\phi)$ is the volume of a sphere of radius $\phi$ and
$\cV_{\ccD}$ is the total volume of projective Hilbert space in $d$
dimensions.  The entropy of a mixture of vectors distributed uniformly
within a sphere of radius $\phi$ is given by Eqs.~(B5)--(B6)
of~\cite{Schack1994a},
\begin{equation}
\HD=-\lambda_0\log\lambda_0
-(1-\lambda_0)\log\!\left({1-\lambda_0\over\ccD-1}\right)
=H_2(\lambda_0)+(1-\lambda_0)\log(\ccD-1)\;,
\label{eq:HDphi}
\end{equation}
where $H_2(\lambda_0)$ is the binary entropy corresponding to the
largest eigenvalue
\begin{equation}
\lambda_0=1-{\ccD-1\over\ccD}\sin^2\!\phi\;.
\label{eq:lambdazero}
\end{equation}

If we group the $\cN$ vectors into groups of radius $\phi$, the
number of vectors per group is
\begin{equation}
\cN_V(\phi)={\cN\over\ND}=\cN(\sin^2\!\phi)^{\ccD-1}\;,
\end{equation}
provided this number is not less than one.  There is a critical
angle, $\phi_{\,b}$, at which there is only one vector per group,
i.e., $(\sin^2\!\phi_{\,b})^{\ccD-1}=1/\cN$.  For
$\phi\ge\phi_{\,b}$, there are $\ND$ groups, each containing
approximately $\cN_V(\phi)$ vectors, but for $\phi\le\phi_{\,b}$,
there are $\cN$ groups, each containing just one vector.  The
information required to specify a group at resolution angle $\phi$
is thus $I(\phi)=\log\cN$ for $\phi\le\phi_{\,b}$ and
$I(\phi)=\log\ND$ for $\phi\ge\phi_{\,b}$.  There is another
critical angle, $\phi_d$, at which there are only two groups,
i.e., $(\sin^2\!\phi_d)^{\ccD-1}=1/2$.  For $\phi\ge\phi_d$, we
can not talk about grouping the vectors into spheres of equal
radius, so we remove these angles $\phi$ from consideration.  Thus
we write the information to specify a group~as
\begin{equation}
I(\phi)=\cases{
    \log\cN\;,&$\phi\le\phi_{\,b}$,\cr
    \log\ND=-(\ccD-1)\log(\sin^2\!\phi)\;,&$\phi_{\,b}\le\phi\le\phi_d$.}
\label{eq:infotwo}
\end{equation}

For $\phi_{\,b}\le\phi\le\phi_d$, we have
$\sin^2\!\phi=2^{-I/(\ccD-1)}=e^{-I\ln2/(\ccD-1)}$, which shows that
there are two important cases in terms of the number of vectors.
If $\log\cN\ll\ccD$ ($\cN\ll2^\ccD$), a situation we refer to as a
{\it sparse\/} collection of random vectors, we have
$I\le\log\cN\ll\ccD$, giving
\begin{equation}
\sin^2\!\phi\approx1-{I\ln2\over\ccD-1}
\qquad\Longrightarrow\qquad
\phi\approx{\pi\over2}-\sqrt{{I\ln2\over\ccD-1}}
\end{equation}
over the entire range $\phi_{\,b}\le\phi\le\phi_d$.  In
particular, we have $\phi_b\approx\pi/2-\sqrt{\ln\cN/(\ccD-1)}$. The
number of groups increases so fast as $\phi$ retreats from $\pi/2$
that for a sparse collection, there is a group for each vector
when the radius $\phi$ is still quite close to $\pi/2$. In
contrast, if $\log\cN\gg\ccD$ ($\cN\gg2^{\ccD}$), which we call a
{\it dense\/} collection of vectors, then
$\phi_{\,b}\approx\sin\phi_{\,b}= 2^{-\log\cN/2(\ccD-1)}\ll1$,
meaning that to get to one vector per group, the radius
$\phi_{\,b}$ must be small.

When we turn to the entropy of the groups, it becomes clear that
there is yet another critical angle, $\phi_c$, the angle at which
the number of vectors per group equals the Hilbert-space
dimension, i.e., $\cN_V(\phi_c)=\ccD$ or
$I(\phi_c)=\log\cN-\log\ccD$. For $\phi\ge\phi_c$, there are
sufficiently many vectors in each group to explore all the
available Hilbert-space dimensions, so the entropy is close
to the entropy $\HD$ of a mixture of vectors distributed uniformly
within a sphere of radius $\phi$ in $\ccD$ dimensions. In contrast,
for $\phi_{\,b}\le\phi\le\phi_c$, the vectors in a group can
explore roughly only $\cN_V(\phi)=\cN(\sin^2\!\phi)^{\ccD-1}=
2^{-I}\cN$ dimensions, thus giving an entropy close to $\HNmax$.
Finally, for $\phi\le\phi_{\,b}$, there is only one vector per
group, so $H=0$.

Our main interest is the relation between $H$ and $I$, so we
eliminate the radius $\phi$ from the above expressions.  The region
$\phi\le\phi_{\,b}$ gives $H=0$ and $I=\log\cN$.  For
$\phi_{\,b}\le\phi\le\phi_c$, i.e., $\log\cN\ge
I\ge\log\cN-\log\ccD$, we have
\begin{equation}
H=\HNmax=
H_2(\lambda)+(1-\lambda)\log\Bigl(2^{-I}\cN-1\Bigr)\;,
\end{equation}
where
\begin{equation}
\lambda=1-{2^{-I}\cN-1\over2^{-I}\cN}2^{-I/(\ccD-1)}
=1-2^{-I/(\ccD-1)}\left(1-{2^{\,I}\over\cN}\right)\;.
\label{eq:lambdadef}
\end{equation}
Finally, for $\phi_c\le\phi\le\phi_d$, i.e., $\log\cN-\log\ccD\ge
I\ge1$, we have $H=\HD$, with
\begin{equation}
\lambda_0=1-{\ccD-1\over\ccD}2^{-I/(\ccD-1)}\;.
\label{eq:lambdazerodef}
\end{equation}
Summarizing, we have
\begin{equation}
H=\cases{
    H_{\cN_V(\phi)}(\phi)
    =H_2(\lambda)+(1-\lambda)\log\Bigl(2^{-I}\cN-1\Bigr)\;,
        &$\log\cN\ge I\ge\log\cN-\log\ccD$,\cr
    \HD=H_2(\lambda_0)+(1-\lambda_0)\log(\ccD-1)\;,&$\log\cN-\log\ccD\ge I\ge 1$.}
\label{eq:tradeofftwo}
\end{equation}
with $\lambda$ and $\lambda_0$ given by Eqs.~({\ref{eq:lambdadef})
and~({\ref{eq:lambdazerodef}).  Equation~(\ref{eq:tradeofftwo}) is
the approximate trade-off relation we are seeking.

The important part of the trade-off relation is the part that is
independent of the number of random vectors, i.e., for $1\le
I\le\log\cN-\log\ccD$.  Notice that to investigate this region, we
need $\cN\gg\ccD$, but we do {\it not\/} need $\cN$ so large that
the random vectors sample generic vectors, which would require at
least $\cN\sim2^{\ccD}$ vectors, i.e., a dense collection. We
emphasize that {\it we do not need a dense collection of vectors
to investigate the important part of the trade-off relation.}

Before going further, it is useful to put the trade-off
relation~(\ref{eq:tradeofftwo}) in other forms, which can be
easily specialized to the case of a sparse collection of vectors.
For the second case, which is the case of interest, we can write
\begin{eqnarray}
\HD&=&
\log\ccD-
{1\over\ccD}
\Bigl(
[\ccD(1-2^{-I/(\ccD-1)})+2^{-I/(\ccD-1)}]\nonumber\\
&\mbox{}&\phantom{=\log\ccD-{1\over\ccD}\Bigl(}
\times\log[\ccD(1-2^{-I/(\ccD-1)})+2^{-I/(\ccD-1)}]
-I2^{-I/(\ccD-1)}\Bigr)\;.
\end{eqnarray}
For a sparse collection of vectors, for which
$I\le\log\cN-\log\ccD\le\log\cN\ll\ccD$, or anytime we have
$I\ll\ccD$, we can approximate this by
\begin{equation}
\HD=\log\ccD-
{1\over\ccD}
\Bigl((1+I\ln2)\log(1+I\ln2)-I\Bigr)\;.
\end{equation}
We can manipulate the first case in Eq.~(\ref{eq:tradeofftwo}) in a
similar way:
\begin{equation}
\HNmax=
\log\cN-I-\lambda\log\!\left({\cN\lambda\over2^{\,I}}\right)
-(1-\lambda)\log\!\left({1-\lambda\over1-2^{\,I}/\cN}\right)\;.
\label{eq:HNmaxtwo}
\end{equation}
The factor $2^{\,I}/\cN$ increases from $1/\ccD$ at
$I=\log\cN-\log\ccD$ to 1 at $I=\log\cN$.  For a sparse collection,
we can approximate $\lambda$ by
\begin{equation}
\lambda=
{2^{\,I}\over\cN}+
{I\ln2\over\ccD-1}\left(1-{2^{\,I}\over\cN}\right)\;.
\end{equation}
The second term is always small.  When the first term dominates,
the second two terms in Eq.~(\ref{eq:HNmaxtwo}) are small.  When
the first term is as small or smaller than the second, the second
two terms in Eq.~(\ref{eq:HNmaxtwo}) are again small.  Thus for a
sparse collection, it is always a good approximation to use
$\HNmax=\log\cN-I$.

The conclusion of these considerations is that for sparse
collections, the trade-off relation~(\ref{eq:tradeofftwo}) is well
approximated by
\begin{equation}
H=\cases{
    \log\cN-I\;,
        &$\log\cN\ge I\ge\log\cN-\log\ccD$,\cr
    \log\ccD-\displaystyle{{1\over\ccD}}
    \Bigl((1+I\ln2)\log(1+I\ln2)-I\Bigr)\;,&$\log\cN-\log\ccD\ge I\ge1$.}
\end{equation}
This is the form of the trade-off relation that we use in
Sec.~\ref{subsubsec:quanthyper}.  When $\ccD$ is large, it is quite
a good approximation for sparse collections of random vectors, certainly
more than adequate given the approximate character of the entire
sphere-grouping model.  These approximate expressions are poorest
at the knee between the two behaviors, which is also where the
approximate treatment of the grouping is at its worst.

\section{Entropy of equal partitions of projective Hilbert space}
\label{app:entropy}

Let $|e_j\rangle$, $j=1,\ldots,\ccD$, be an orthonormal basis for a
$\ccD$-dimensional Hilbert space, and let
\begin{equation}
\hat{P}_+=\sum_{j=1}^n|e_j\rangle\langle e_j|
\end{equation}
be the projector onto the subspace $S_+$ spanned by the first $n$
vectors,
\begin{equation}
\hat{P}_-=\sum_{j=n+1}^{n+m}|e_j\rangle\langle e_j|
\end{equation}
be the projector onto the subspace $S_-$ spanned by the next $m$
vectors, and $\hat{P}_0=\one-\hat{P}_+-\hat{P}_-$ be the projector onto the subspace
$S_0$ spanned by the remaining $\ccD-n-m$ vectors.  An arbitrary
normalized vector can be expanded uniquely as
\begin{equation}
|\psi\rangle=
\cos\xi(\cos\theta|\chi\rangle+\sin\theta|\eta\rangle)
+\sin\xi|\phi\rangle\;,
\end{equation}
where $|\chi\rangle\in S_+$, $|\eta\rangle\in S_-$, and $|\phi\rangle\in
S_0$ are normalized vectors.  The angle $\xi$ is the Hilbert-space angle
between $|\psi\rangle$ and the span of $S_+$ and $S_-$, and $\theta$ is
the Hilbert-space angle between the projection of $|\psi\rangle$ into the
span of $S_+$ and $S_-$, i.e., $(\hat{P}_++\hat{P}_-)|\psi\rangle$, and the subspace
$S_+$.

We are interested in the density operator formed from all pure states
whose projection into the span of $S_+$ and $S_-$ is closer to $S_+$ than
an angle $\Theta$,
\begin{equation}
\r=\sN\int_{\theta\le\Theta}\d\Sall\,|\psi\rangle\langle\psi|\;,
\end{equation}
where $\sN$ is a normalization factor.  Here and throughout $\d\cS_j$
denotes the standard integration measure on the $j$-sphere, and
$\cS_j=\int \d\cS_j$ is the volume of the $j$-sphere.  This region of
states is the analogue of the intersection in three real dimensions of a
wedge of opening angle $2\Theta$ with the unit sphere.  It is clear that
$\r$ is invariant under unitary transformations that are block-diagonal in
the three subspaces, which implies that $\r$ has the form
\begin{equation}
\r=\lambda_+ \hat{P}_+ + \lambda_- \hat{P}_- +\lambda_0 \hat{P}_0\;.
\end{equation}
Our job is to determine the three eigenvalues, $\lambda_\pm$ and
$\lambda_0$, which satisfy
\begin{equation}
n\lambda_+ +m\lambda_- + (\ccD-n-m)\lambda_0=1\;.
\end{equation}
It turns out that $\lambda_0=1/\ccD$, as we show below, so we have
\begin{equation}
\lambda_-=
{1\over\ccD}
\left(1+{n\over m}(1-\ccD\lambda_+)\right)
\;.
\label{eq:lambdaminus}
\end{equation}

A small change in $|\psi\rangle$ can be written as
\begin{eqnarray}
|\d\psi\rangle&=&
\d\xi\Bigl(-\sin\xi(\cos\theta|\chi\rangle+\sin\theta|\eta\rangle)
+\cos\xi|\phi\rangle\Bigr)+\sin\xi|\d\phi\rangle\nonumber\\
&\mbox{}&\phantom{\d\xi}
+\cos\xi\Bigl(\d\theta(-\sin\theta|\chi\rangle+\cos\theta|\eta\rangle)
+\cos\theta|\d\chi\rangle+\sin\theta|\d\eta\rangle\Bigr)\;.
\end{eqnarray}
This gives a line element on normalized vectors,
\begin{equation}
\d s^2=\langle \d\psi|\d\psi\rangle=
\d\xi^2+\sin^2\!\xi\langle \d\phi|\d\phi\rangle
+\cos^2\!\xi
\Bigl(\d\theta^2+\cos^2\!\theta\langle \d\chi|\d\chi\rangle
+\sin^2\!\theta\langle \d\eta|\d\eta\rangle\Bigr)\;,
\end{equation}
and a corresponding volume element on the ($2\ccD-1$)-sphere of normalized
vectors,
\begin{eqnarray}
\d\Sall&=&
\sin^{2(\ccD-n-m)-1}\!\xi\cos^{2(n+m)-1}\!\xi\,\d\xi\nonumber\\
&\mbox{}&\phantom{(}\times
\cos^{2n-1}\!\theta\sin^{2m-1}\!\theta\,\d\theta\,
\d\Szero\,\d\Splus\,\d\Sminus\;.
\end{eqnarray}
Normalizing the density operator gives
\begin{eqnarray}
1={\rm tr}(\r)&=&
\sN\int_{\theta\le\Theta}\d\Sall\nonumber\\
&=&\sN\Szero\Splus\Sminus
\int_0^{\pi/2}\d\xi\,\sin^{2(\ccD-n-m)-1}\!\xi\cos^{2(n+m)-1}\!\xi\nonumber\\
&\mbox{}&\phantom{=\int}\times
\int_0^\Theta \d\theta\,\cos^{2n-1}\!\theta\sin^{2m-1}\!\theta\;.
\label{eq:normalization}
\end{eqnarray}

We first verify that $\lambda_0=1/\ccD$.  Letting $|e_0\rangle$ be
any normalized vector in $S_0$, we have
\begin{eqnarray}
\lambda_0=\langle e_0|\r|e_0\rangle
&=&\sN\int_{\theta\le\Theta}\d\Sall|\langle e_0|\psi\rangle|^2\nonumber\\
&=&\sN\Splus\Sminus
\int_0^{\pi/2}\d\xi\,\sin^{2(\ccD-n-m)+1}\!\xi\cos^{2(n+m)-1}\!\xi\nonumber\\
&&\phantom{=\int}\times
\int_0^\Theta \d\theta\,\cos^{2n-1}\!\theta\sin^{2m-1}\!\theta
\int \d\Szero\,|\langle e_0|\phi\rangle|^2\;.
\end{eqnarray}
Using
\begin{equation}
\int \d\Szero\,|\langle e_0|\phi\rangle|^2={\Szero\over\ccD-n-m}
\end{equation}
and the expression for the normalization constant from
Eq.~(\ref{eq:normalization}) and changing integration variable
to $u=\sin^2\!\xi$, we get
\begin{eqnarray}
\lambda_0
&=&{1\over\ccD-n-m}
{\displaystyle{\int_0^1 \d u\,u^{\ccD-n-m}(1-u)^{n+m-1}}
\over
\displaystyle{\int_0^1 \d u\,u^{\ccD-n-m-1}(1-u)^{n+m-1}}
}\nonumber\\
&=&{1\over\ccD-n-m}
{\Gamma(\ccD-n-m+1)\Gamma(n+m)/\Gamma(\ccD+1)
\over
\Gamma(\ccD-n-m)\Gamma(n+m)/\Gamma(\ccD)}\nonumber\\
&=&{1\over\ccD}\;.
\end{eqnarray}

Similarly, to find $\lambda_+$, we let $|e_+\rangle$ be any
normalized vector in $S_+$ and write
\begin{eqnarray}
\lambda_+=\langle e_+|\r|e_+\rangle
&=&\sN\int_{\theta\le\Theta}\d\Sall
|\langle e_+|\psi\rangle|^2\nonumber\\
&=&\sN\Szero\Sminus
\int_0^{\pi/2}\d\xi\,\sin^{2(\ccD-n-m)-1}\!\xi\cos^{2(n+m)+1}\!\xi\nonumber\\
&&\phantom{=\int}\times
\int_0^\Theta \d\theta\,\cos^{2n+1}\!\theta\sin^{2m-1}\!\theta
\int \d\Splus\,|\langle e_+|\chi\rangle|^2\;.
\end{eqnarray}
Using
\begin{equation}
\int \d\Splus\,|\langle e_+|\chi\rangle|^2={\Splus\over n}
\end{equation}
and the expression for the normalization constant and changing
integration variables to $u=\sin^2\!\xi$ and $v=\sin^2\!\theta$,
we get
\begin{eqnarray}
\lambda_+
&=&{1\over n}
{
\displaystyle{\int_0^1 \d u\,u^{\ccD-n-m-1}(1-u)^{n+m}}
\over
\displaystyle{\int_0^1 \d u\,u^{\ccD-n-m-1}(1-u)^{n+m-1}}
}
{
\displaystyle{\int_0^{\sin^2\!\Theta}\d v\,v^{m-1}(1-v)^n}
\over
\displaystyle{\int_0^{\sin^2\!\Theta}\d v\,v^{m-1}(1-v)^{n-1}}
}\nonumber\\
&=&{1\over n}
{
\displaystyle{{\Gamma(\ccD-n-m)\Gamma(n+m+1)/\Gamma(\ccD+1)\over
\Gamma(\ccD-n-m)\Gamma(n+m)/\Gamma(\ccD)}}
}
{
\displaystyle{\int_0^{\sin^2\!\Theta}\d v\,v^{m-1}(1-v)^n}
\over
\displaystyle{\int_0^{\sin^2\!\Theta}\d v\,v^{m-1}(1-v)^{n-1}}
}\nonumber\\
&=&{n+m\over n\ccD}
{
\displaystyle{\int_0^{\sin^2\!\Theta}\d v\,v^{m-1}(1-v)^n} \over
\displaystyle{\int_0^{\sin^2\!\Theta}\d v\,v^{m-1}(1-v)^{n-1}}
}
\;.
\end{eqnarray}

We now specialize to the case of interest, $n=m$ and $\Theta=\pi/4$, so
that $\r$ is constructed from pure states occupying one of two halves
of Hilbert space:
\begin{equation}
\lambda_+=
{2\over\ccD}
{
\displaystyle{\int_0^{1/2}\d v\,v^{n-1}(1-v)^n}
\over
\displaystyle{\int_0^{1/2}\d v\,v^{n-1}(1-v)^{n-1}}
}
\;.
\label{eq:lambdaplus}
\end{equation}
The integrals can be evaluated as
\begin{eqnarray}
\int_0^{1/2}\d v\,v^{n-1}(1-v)^n&=&
{n!(n-1)!\over2(2n)!}
\left(1+{\Gamma(n+1/2)\over\sqrt\pi\,n!}\right)\;,\nonumber\\
\int_0^{1/2}\d v\,v^{n-1}(1-v)^{n-1}&=&
{[(n-1)!]^2\over2(2n-1)!}\;.
\end{eqnarray}
Plugging these results into Eqs.~(\ref{eq:lambdaplus}) and
(\ref{eq:lambdaminus}), we get
\begin{equation}
\lambda_\pm=
{1\over\ccD}\left(1\pm{\Gamma(n+1/2)\over\sqrt\pi\,n!}\right)
={1\over\ccD}\left(1\pm{(2n)!\over2^{2n}(n!)^2}\right)\;.
\label{eq:lambdapm}
\end{equation}
When $n=1$, we get $\lambda_+=3/2\ccD$ and $\lambda_-=1/2\ccD$, and when
$n=2$, $\lambda_+=11/8\ccD$ and $\lambda_-=5/8\ccD$.  For large $n$ (and
$\ccD$), we can use Stirling's formula to write
\begin{equation}
\lambda_\pm\approx{1\over\ccD}
\left(1\pm{1\over\sqrt{\pi n}}\right)\;.
\label{eq:lambdapmapprox}
\end{equation}

The von Neumann entropy of $\r$ can be put in the form
\begin{eqnarray}
H&=&
-n\lambda_+\log\lambda_+ -n\lambda_-\log\lambda_-
-(\ccD-2n)\lambda_0\log\lambda_0\nonumber\\
&=&
\log\ccD-{2n\over\ccD}\Bigl(1-H_2(\ccD\lambda_+/2)\Bigr)\;.
\label{eq:HnD}
\end{eqnarray}
For fixed $\ccD$, this is a decreasing function of $n$.  For large
$n$ (and $\ccD$), we can use Eq.~(\ref{eq:lambdapmapprox}) to write
$H_2(\ccD\lambda_+/2)\approx1-1/2\pi n\ln2$ and
\begin{equation}
H\approx\log\ccD-{1\over\pi\ccD\ln2}=\log\ccD-{0.46\over\ccD}\;.
\label{eq:HnDapprox}
\end{equation}

\end{document}